\documentclass[showpacs,showkeys,twocolumn,pra,superscriptaddress]{revtex4-1}%
\pdfoutput=1
\usepackage{graphics,amsmath,amsfonts,amscd,revsymb,latexsym,
enumerate,multirow,epsfig}
\usepackage{times}
\usepackage{mathtools}
\usepackage[caption=false]{subfig}
\usepackage{amssymb}
\usepackage{graphicx}%
\usepackage{bbm}
\providecommand{\U}[1]{\protect\rule{.1in}{.1in}}
\providecommand{\U}[1]{\protect\rule{.1in}{.1in}}

\begin{document}
\title{Fermion bag approach to Hamiltonian lattice field theories in continuous time}

\author{Emilie Huffman}
\author{Shailesh Chandrasekharan}
\affiliation{Duke University, Durham, North Carolina 27708, USA}

\begin{abstract}
We extend the idea of fermion bags to Hamiltonian lattice field theories in the continuous time formulation. Using a class of models we argue that the temperature is a parameter that splits the fermion dynamics into small spatial regions that can be used to identify fermion bags. Using this idea we construct a continuous time quantum Monte Carlo algorithm and compute critical exponents in the $3d$ Ising Gross-Neveu universality class using a single flavor of massless Hamiltonian staggered fermions. We find $\eta=0.54(6)$ and $\nu=0.88(2)$ using lattices up to $N=2304$ sites. We argue that even sizes up to $N=10,000$ sites should be accessible with supercomputers available today.
\end{abstract}
\maketitle
 
\section{Introduction}
Quantum Monte Carlo (QMC) methods of studying strongly correlated fermion systems are known to be notoriously difficult \cite{Troyer:2004ge}. Even if sign problems can be solved it is difficult to study large system sizes close to critical points, especially when the system contains long range fermionic correlations. Many strongly interacting quantum critical points were predicted long ago in $2+1$ dimensions in the presence of massless Dirac fermions \cite{PhysRevLett.62.1433,Rosenstein199159}, but 
their properties have not yet been determined accurately using quantum Monte Carlo methods. Due to developments in condensed matter physics related to the physics of graphene and the associated developments in topological insulators the field has become interesting again and there is new impetus to study the critical points better \cite{PhysRevLett.97.146401,PhysRevD.86.105007,Wellegehausen:2017goy,You:2017ltx}.

Studies based on the Lagrangian formulation on space-time lattices use the Hybrid Monte Carlo (HMC) algorithm \cite{Hands199329,Karkkainen94,PhysRevD.53.4616,Stavros07}. Although it is expected to have better scaling properties with system size compared to other fermion algorithms, it encounters singularities in the presence of massless fermions, especially near strongly interacting quantum critical points. In order to avoid such singularities, studies include a fermion mass. The presence of two infrared scales, in the form of a fermion mass and a finite lattice size, makes accurately extracting the critical exponents difficult. Ways to circumvent these problems could be very helpful.

Lagrangian formulations have other limitations as well. Ultra-local actions on space-time lattices can create extra doubling of fermion degrees of freedom due to time discretization. Along with chiral symmetry some internal flavor symmetries may also be lost. For example, the semi-metal insulator phase transition in graphene that was studied recently using the Lagrangian formulation with staggered fermions, breaks the important $SU(2)$ spin symmetry \cite{PhysRevB.79.165425}. Recently, Lagrangian formulations of Dirac fermions in $2+1$ dimensions have begun to use overlap or domain wall fermions  \cite{PhysRevD.94.065026,Hands:2016foa,Hands:2017hhk}. While these formulations preserve many symmetries of continuum Dirac fermions, they are computationally much more expensive, especially near strongly coupled quantum critical points.

We can circumvent some of the limitations of Lagrangian formulations by constructing the partition function starting from a lattice Hamiltonian. Since we can eliminate time discretization errors we can avoid an extra fermion doubling and preserve more symmetries \cite{PhysRevB.72.035122,RevModPhys.83.349,Wang:2015rga}. Also, unlike the HMC approach the auxiliary field Monte Carlo (AFMC) methods used in the Hamiltonian formulation can also work with exactly massless fermions without encountering singularities \cite{PhysRevD.24.2278,Bercx:2017pit}. In principle the time to perform a single sweep in AFMC can be reduced to scale as $\beta N^3$ where $N$ is the number of spatial sites and $\beta$ is the inverse temperature. However, there can be bottlenecks due to numerical instabilities on large lattices. Several recent studies of semi-metal-insulator phase transitions in $2+1$ Dirac systems have emerged recently using this approach \cite{1367-2630-16-10-103008,1367-2630-17-8-085003,Hesselmann:2016tvh}, and the largest lattices explored are roughly of the order of $N=2500$ on honeycomb lattices and $N=1600$ on square lattices \cite{,PhysRevX.6.011029}. Calculations in the continuous time limit involve much smaller sizes. Recently, the HMC algorithm has also been applied to Hamiltonian formulations \cite{PhysRevLett.111.056801,Korner:2017qhf,Beyl:2017kwp}, but the problems related to singularities mentioned above continue to be a bottleneck.

Recently a new idea called the fermion bag approach, has been used to accelerate fermion algorithms \cite{PhysRevD.82.025007,Chandrasekharan2013}. The idea was originally formulated within the Lagrangian formulation and has allowed us to study large lattices with exactly massless Dirac fermions and accurately extract critical exponents at some of the quantum critical points in $2+1$ dimensions \cite{PhysRevLett.108.140404,PhysRevD.88.021701}. In this work we extend the idea to Hamiltonian formulations in continuous time. Using it we are able to study lattices containing up to $N=10,000$ sites without encountering numerical instabilities. Although computing quantities close to quantum critical points on such large lattices still requires supercomputers, we are able to study square lattices with up to $N=2304$ sites on small computer clusters.

\section{Idea of Fermion Bags}

The idea of fermion bags is based on the intuition that it should be possible to write a fermionic partition function as a sum over weights of configurations where each configuration weight is obtained as a product of weights of smaller configurations. This is accomplished by dividing the fermion degrees of freedom of the entire system into many smaller entangled regions (or fermion bags) that are essentially independent of each other \cite{PhysRevD.82.025007}. The fermion bag weight is obtained by summing over all quantum fluctuations within the bag. If this weight is positive an efficient Monte Carlo algorithm could be designed. The idea of fermion bags is an extension of the meron cluster approach  \cite{PhysRevLett.83.3116}.

While the idea of fermion bags is widely applicable there is no unique recipe to identify the bags for a given model. One guiding principle is that weights of fermion bags must be positive which is not always guaranteed. One can also use efficiency of Monte Carlo sampling as the other guiding principle. If the fermion bags can identify the entanglement that arises naturally from the underlying physics and fermion bag weights remain positive, then the Monte Carlo sampling usually becomes efficient. For example, fermion bags can be identified differently at strong couplings as compared to weak couplings. At weak couplings Feynman diagrams suggest a natural choice for the fermion bags and then the approach is identical to the determinantal diagrammatic Monte Carlo methods \cite{PhysRevE.74.036701,PhysRevLett.101.090402,PhysRevB.76.035116}. But such an identification leads to inefficient Monte Carlo sampling at stronger couplings since the entanglement of the fermion degrees of freedom changes. Efficiency can be improved by combining weak and strong coupling fermion bags at intermediate couplings.

Recently we discovered that the idea of fermion bags can be useful even if a fermion bag becomes entangled with the rest of the system. We realized that this entanglement can be stored in the form of a large matrix. If this can be computed and stored we can perform fast updates of fermion bags. This extension of the fermion bag idea is similar to the idea of local factorization of the determinant proposed recently  \cite{PhysRevD.95.034503}. In our case it has allowed us to study $60^3$ lattices near a quantum critical point with exactly massless fermions for the first time \cite{PhysRevD.93.081701}. In this work we argue that a similar idea should be applicable for Hamiltonian lattice fermions.

In order to illustrate how the idea of fermion bags can be extended to Hamiltonian formulations in continuous time, in this work we focus on those that can be written as $H = \sum_{x,d} H_{x,d}$ where
\begin{equation}
H_{x,d} \ =\ -\omega_{\langle x,d\rangle}\ 
\mathrm{e}^{2\alpha_{\langle x,d\rangle}\ 
\sum_{a=1}^{N_f} \big({c^a_x}^\dagger c^a_{x+\hat{d}} + {c^a_{x+\hat{d}}}^\dagger c^a_x\big)}.
\label{fact}
\end{equation}
Here $x$ is a spatial lattice site, $\hat{d}$ labels the directions such that $\langle x,d\rangle$ labels a unique nearest neighbor bond. The operators ${c^a_x}^\dagger$ and $c^a_x$ are fermionic creation and annihilation operators at the site $x$ with a flavor $a=1,2..,N_f$. The couplings of the model are defined through the real constants $\delta_{\langle x,d\rangle} > 0$ and $\alpha_{\langle x,d\rangle}$. In the discussions below we focus on the $N_f=1$ model on a two dimensional square lattice with periodic boundary conditions and $L$ sites in each direction with $N=L^2$. However, they can be extended to any value of $N_f$ and all bi-partite lattice models where the sites connected to the bond $\langle x,d\rangle$ lie on different sub-lattices.

Although the Hamiltonians we consider are unconventional they contain rich physics. We have designed them so that the idea of fermion bags is applicable \cite{RevCondMat}. For a fixed $N_f$ they are invariant under an $O(2N_f)$ flavor symmetry in addition to the usual lattice symmetries, some of which may be broken spontaneously at quantum critical points  \cite{PhysRevX.6.041049}. When $N_f=1$ our model is equivalent (up to an constant) to the $t-V$ model,
\begin{equation}
H_{x,d} \ =\ - t\eta_{x,d}\left(c_x^\dagger c_{x+\hat{d}} + c_{x+\hat{d}}^\dagger c_x\right)
 - V \Phi_x \Phi_{x+\hat{d}},
\label{tvmodel}
\end{equation}
when $V > 0$. Here we define $\Phi_x =(-1)^{x_1+x_2}(c^\dagger_x c_x - 1/2)$, assuming a lattice site with coordinates $x=(x_1,x_2)$. The equivalence requires that we set $\omega_{\langle x,d\rangle }=t^2/(V (1-\left(V/2t\right)^2)$, and $\alpha_{\langle x,d\rangle } = \alpha \eta_{x,d}$ where $\cosh2\alpha=(1+\left(V/2t\right)^2)/(1-\left(V/2t\right)^2)$, and $\sinh2\alpha (V/t)/(1-\left(V/2t\right)^2)$ \cite{PhysRevB.93.155117}. If we define $\eta_{\langle x,1\rangle} = 1$ and $\eta_{\langle x,2\rangle} = (-1)^{x_1}$, the model describes interacting two dimensional massless Hamiltonian staggered fermions \cite{PhysRevD.16.3031}.

\begin{figure*}[ht]
\minipage{0.3\textwidth}
  \includegraphics[width=\linewidth]{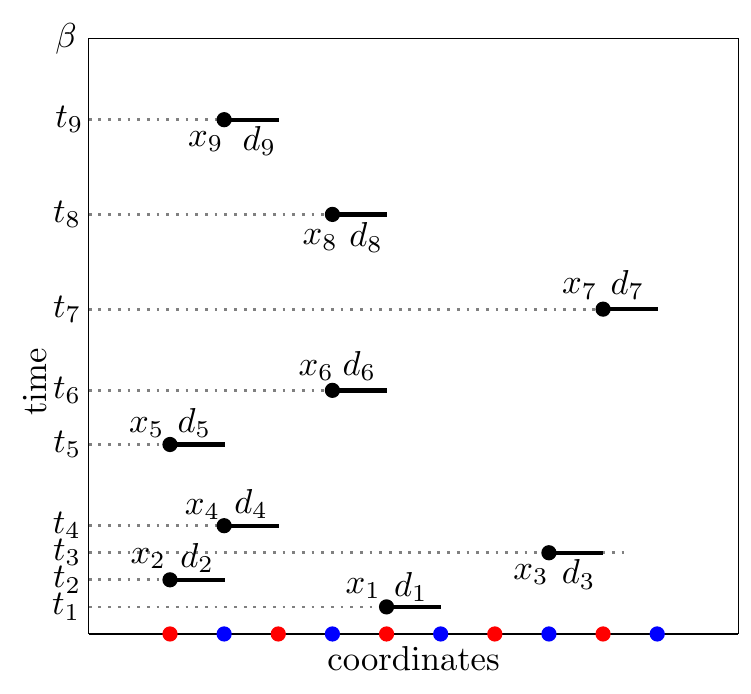}
  \caption{An example configuration. The horizontal axis labels the spatial sites, the vertical axis is imaginary time.}\label{exp}
\endminipage\hfill
\minipage{0.29\textwidth}
  \includegraphics[width=\linewidth]{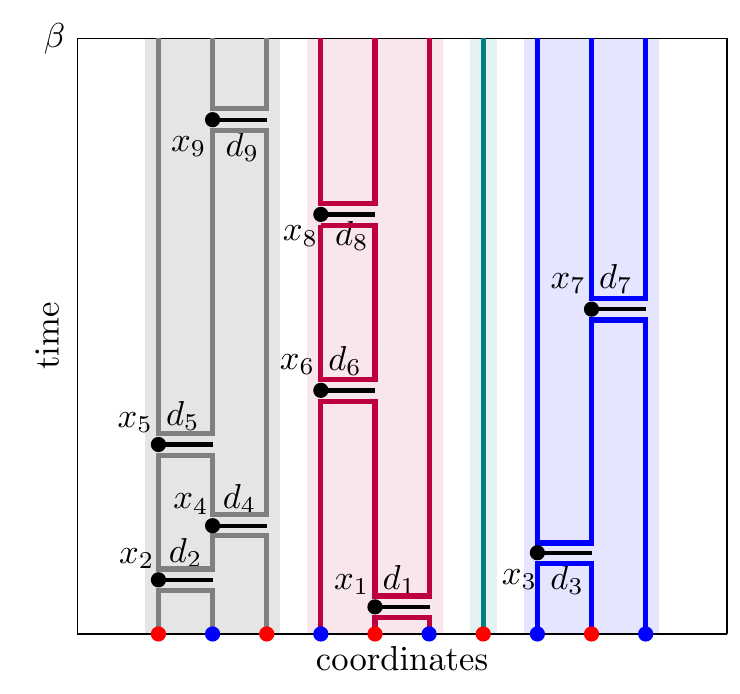}
  \caption{The bonds in this configuration form four fermion bags between $t=0$ and $t=\beta$.}\label{fb}
\endminipage\hfill
\minipage{0.37\textwidth}%
  \includegraphics[width=\linewidth]{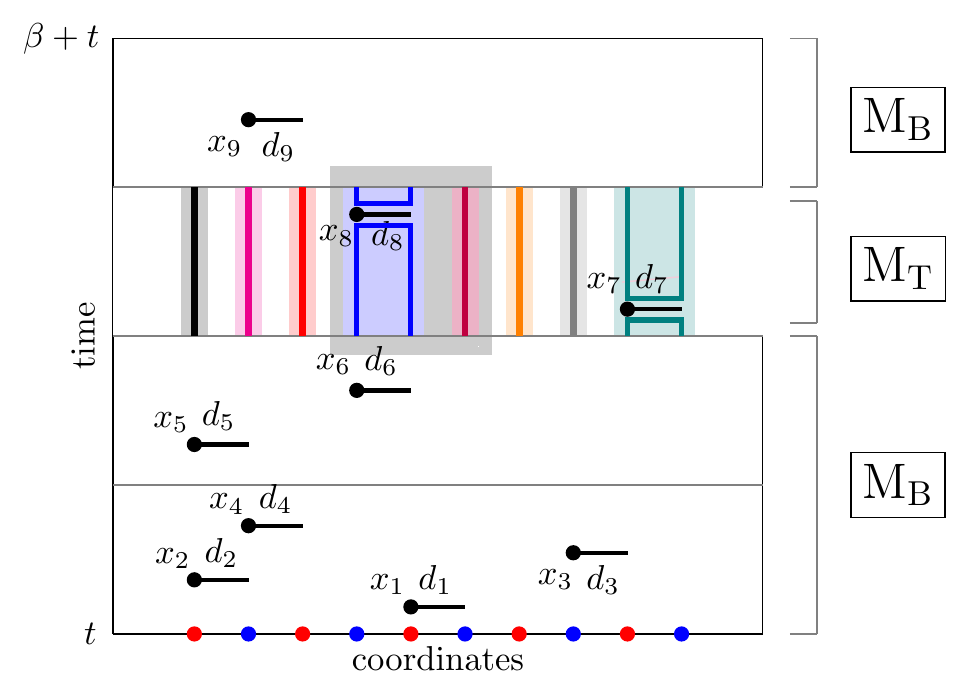}
  \caption{Timeslices are added and $M_T$ and $M_B$ regions defined. Fermion bags are highlighted in the $M_T$ region, and the current update block is shaded.}\label{regions}
\endminipage
\end{figure*}

Using the well known CT-INT expansion of the partition function \cite{PhysRevB.72.035122,RevModPhys.83.349,Wang:2015rga} we can write
\begin{equation}
Z\ =\ \sum_k\ \int \ [dt]
\sum_{[\langle x,d\rangle]}\ {\rm Tr}
\Big(H_{x_k,d_k}\ ...\ H_{x_2,d_2}\ H_{x_1,d_1}\Big),
\label{partition}
\end{equation}
where there are $k$ insertions of the bond Hamiltonian $H_{x,d}$ inside the trace at times $t_1\leq t_2 \leq ...\leq t_k$. The symbol $[dt]$ represents the $k$ time-ordered integrals and $[\langle x,d\rangle] = \{\langle x_1,d_1\rangle,\langle x_2,d_2\rangle,...\langle x_k,d_k\rangle\}$ represents the configuration of bonds at different times. Since a configuration of bonds also requires the information of the times where the bonds are inserted we label the configuration as $[x,d,t]$. An illustration of the bond configuration is shown in Fig.~\ref{exp}. Each bond represents the operator $H_{x,d}$ that is present inside the trace in (\ref{partition}). It can be shown that the traces that appear in (\ref{partition}) are always positive \cite{Huffman:2013mla,PhysRevB.91.241117}.

We can imagine $H_{\langle x,d\rangle}$ as creating a quantum entanglement between the fermions at $x$ and $x+\hat{d}$. Thus, all spatial sites connected by bonds to each other at various times become entangled with each other. Such a group of entangled sites can be defined as a fermion bag. For the bond configuration in Fig.~\ref{exp} we identify four fermion bags as shown in Fig.~\ref{fb}. When two bonds $\langle x,d\rangle$ and $\langle x',d'\rangle$ do not share a site between them the bond Hamiltonians commute, i.e.,
$\left[H_{\langle x,d\rangle}, H_{\langle x',d'\rangle}\right] = 0$. This implies that the weight of the bond configuration can be written as a product of weights fermion bags.

Since the space-time density of bonds is a physical quantity related to the energy density of the system \cite{PhysRevB.93.155117}, for every coupling $V$ we expect a fixed density of bonds. This implies that we can use the temperature as a parameter to control the size of fermion bags. At high temperatures we will have fewer bonds and many small fermion bags. Note that lattice sites that are not connected to any bonds form their own fermion bag. As the temperature is lowered fermion bags will begin to merge to form a single large fermion bag. At very low temperatures there will only be a few isolated small fermion bags. This suggests that at some optimal temperature the fermion bags may efficiently break up the system into smaller regions that do not depend on the system size. Even at low temperatures, we may be able to divide the imaginary time axis into many time-slices and update a single time-slice efficiently. This is illustrated in Fig.~\ref{regions}, where the imaginary time extent is divided into four-time slices and in the shaded time-slice there are eight fermion bags, instead of the four shown in Fig.~\ref{fb}.

In order to test if the maximum fermion bag size remains independent of the lattice size even for large lattices we have studied the $t-V$ model (\ref{tvmodel}) on a square lattice near its critical point. Taking $\beta=4.0$ we divided the imagninary time direction into $16$ time-slices and studied the fermion bag size as a function of the lattice size. For equilibrated configurations of $L=48, 64$ and $100$, the average maximum fermion bag size within a time slice was about $30$ independent of $L$. Further tests suggests that the optimal temperature is roughly $0.25$. Since bond insertions in different fermion bags commute with each other, we can efficiently update fermion bags in space-time blocks (shown as a box in the shaded time slice in Fig.~\ref{regions})) involving $30$ to $60$ spatial sites within each time slice. During this update the effects of the bonds outside this block is taken into account through the fixed $N\times N$ matrix as we discuss in the next section.

\section{Algorithm and Updates}

We now discuss our Monte Carlo algorithm to calculate the correlation observable 
\begin{equation}
\langle C\rangle = {\rm Tr}\left(\Phi_{(0,0)} \Phi_{(L/2,0)} e^{-\beta H}\right)/{\rm Tr}\left(e^{-\beta H}\right).
\label{obs}
\end{equation}
to illustrate the advantages of the fermion bag approach. This observable is used in the next section to study the quantum critical behavior of the $t-V$ model. In our algorithm we generate configurations $([x,d,t];t_0)$ in
two sectors: the partition function sector ($n=0$) with weight $\Omega_0([x,d,t];t_0)$ and the observable sector ($n=1$) with weight $f\Omega_1([x,d,t];t_0)$ where
\begin{equation}
\begin{aligned}
\Omega_n([x,d,t];t_0) &= {\rm Tr}\left[H_{x_k, d_k} ...C_n... H_{x_2,d_2} H_{x_1, d_1}\right]
\end{aligned}
\end{equation}
Here $0 \leq t_0 \leq \beta$ is a time where the operator $C_n$ is introduced. In the partition function sector $C_0 = I$ (the identity operator) and in the observable sector $C_1 = \Phi_{(0,0)} \Phi_{(L/2,0)}$. The factor $f > 0$ is chosen so that the two sectors can be sampled with roughly equal probabilities. We record the number
 \begin{equation}
 {\cal N}\ =\ \frac{\Omega_1([x,b,t];t_0)}{\Omega_0([x,b,t];t_0) + f\Omega_1([x,b,t];t_0)}
 \label{nobs}
 \end{equation}
for each configuration generated. It is easy to prove that $\langle C \rangle = \langle {\cal N}\rangle/(1-f\langle {\cal N}\rangle$.

We use four different updates to generate the configurations $([x,d,t];t_0)$ in the two sectors: (1) \textit{Sector-update:} We flip the sector $n \rightarrow 1-n$ while keeping $([x,d,t];t_0)$ fixed. This update is time consuming and will be explained further below. (2) \textit{Move-update:} Since bond insertions commute with each other when they do not share a lattice site we can move all the bonds in time as long as two non-commuting operator insertions do not cross each other. We try to move roughly the same number of bonds moves as there are bonds in an equilibrated configuration. During this step $t_0$ and $n$ remain fixed. (3) \textit{Time-update:} $t_0 \leftrightarrow t_0'$ while keeping the bond configuration $[x,d,t]$. We perform this update only in the $n=0$ sector where it is trivial. (4) \textit{Bond-update:} This is the most time consuming update where we attempt to change the entire bond configuration $[x,d,t] \leftrightarrow [x',d',t']$ while keeping $t_0$ and $n$ fixed. We perform exactly one bond update per sweep since it is very expensive. For the other updates we perform a fixed number of each per sweep depending on the lattice size.

The \textit{sector-update} and the \textit{bond-update} are the two most time-intensive updates since we need to compute the ratio
$R = \Omega_n\left([x,b,t];t_0\right)/\Omega_n'\left([x',b',t'];t_0\right)$ to calculate the transition probabilities in the Metropolis accept/reject step. Since the sector update is a special case of the bond update we only focus on the details of the bond updates. Using the BSS formula \cite{inbook} we can show
\begin{equation}
\Omega_n\left([x,t,b];t_0\right)=\det\left(\mathbbm{1}_N+B_{x_k,d_k}...O_n...B_{x_2,d_2} B_{x_1,d_1}\right),
\end{equation}
where $\mathbbm{1}_N$, $B_{x_i,d_i}$ and $O_n$ are all $N\times N$ matrices with rows and columns labeled by spatial lattice sites. The matrix $\mathbbm{1}_N$ is the identity matrix, while $B_{x_i,d_i}$ is the identity matrix except in a $2\times 2$ block labeled by the rows and columns of the sites that touch the bond $\langle x_i,d_i\rangle$. Within this block, $B_{x_i,d_i}$ takes the form 
\begin{equation}
{\cal B}_{x,d} = \left(\begin{array}{cc} \cosh2\alpha & \eta_{x,d}\sinh2\alpha \\ \eta_{x,d}\sinh2\alpha & \cosh2\alpha \end{array}\right).
\end{equation}
Finally, the matrix $O_n$ depends on the sector $n$ and is given by $O_0=\mathbbm{1}_N$ and $(O_1)_{x,y}=\delta_{x,y}- 2\delta_{x,(0,0)} - 2\delta_{x,(L/2,0)}$.

\begin{figure}[t]
\includegraphics[width=0.4\textwidth]{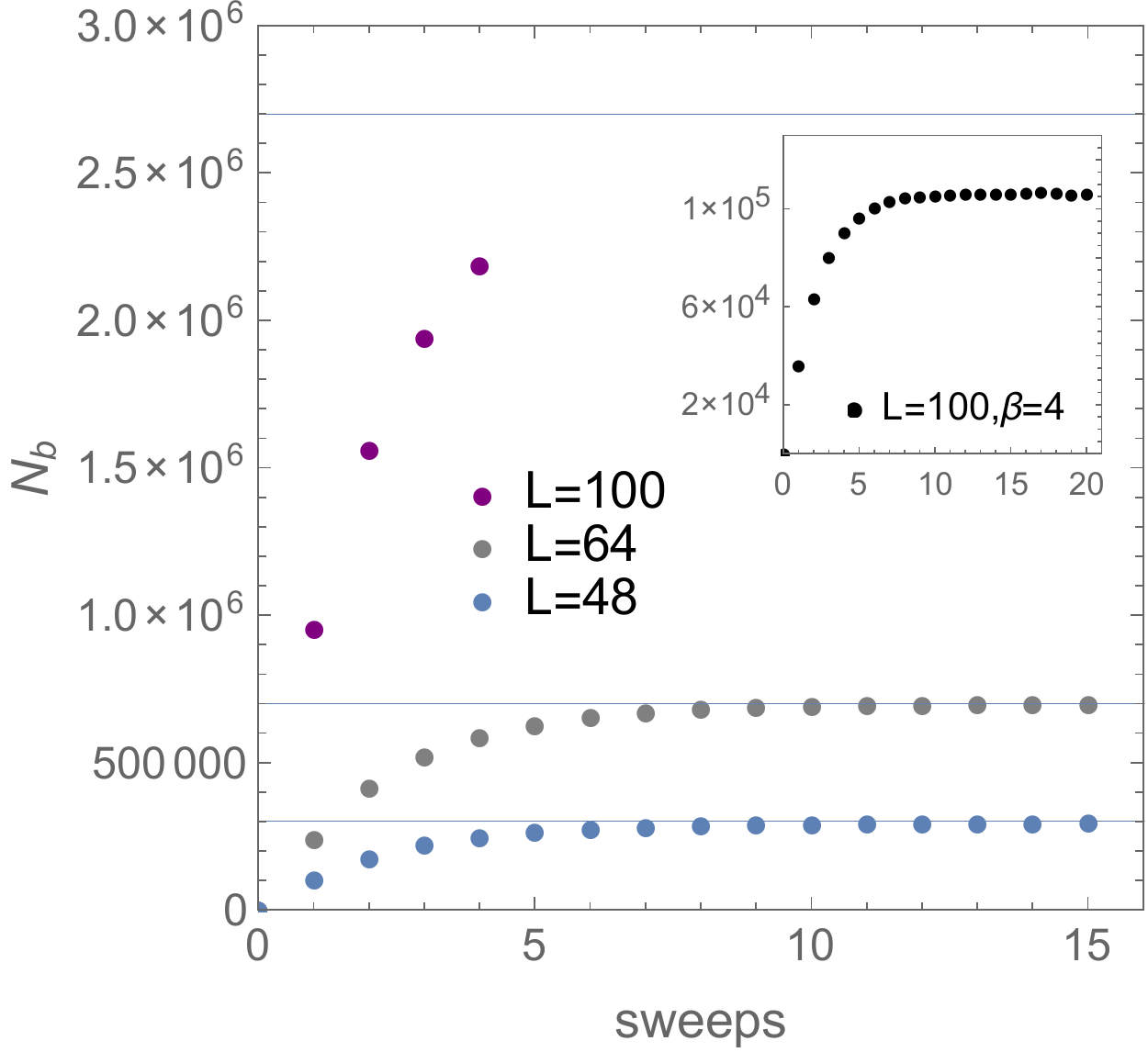}
\caption{Plot showing $\beta=L$ equilibration of the total number of bonds $N_b$ in a bond configuration starting from zero, as a function of Monte Carlo sweeps. The horizontal lines show the expected equilibrated values. The time for a single bond-update on a single core are approximately 30 days for $L=100$, 30 hours for $L=64$, and 4 hours for $L=48$. Inset shows equilibration at $L=100$, $\beta=4$.}
\label{sweepequil}
\end{figure}

Before we begin the bond update we divide the configuration space into time-slices of width $0.25$ with $t_0$ chosen to be at the beginning of the first time slice. We then update bonds within each time-slice sequentially. During the update of a time-slice we define two $N\times N$ matrices: the \textit{background matrix} $M_B$ (which is a product of all of the $B_{x,d}$ matrices outside the selected time-slice and $O_n$), and the \textit{time-slice matrix} $M_T$, which is the product of all the $B_{x,d}$ matrices within the time-slice being updated. Figure~\ref{regions} shows what contributes to $M_B$ and $M_T$. When the configuration of bonds within the time-slice is changed then only $M_T$ changes to $M'_T$. The ratio $R$ is given by
\begin{equation}
R \ =\ \frac{\det(\mathbbm{1}_N + M_B M_T')}
{\det(\mathbbm{1}_N + M_B M_T)} \ =\ \det\left(\mathbbm{1}_N + G_B\Delta \right),
\label{largem}
\end{equation}
where we have defined two new $N\times N$ matrices $G_B = \left(\mathbbm{1}_N+M_B M_T\right)^{-1} M_B M_T$ and $\Delta = \left(M_T^{-1} M'_T - \mathbbm{1}_N\right)$. Since the bond matrices $B_{x,d}$ in different fermion bags commute, it is easy to verify that $\Delta$ is non-zero only within a block which contains spatial sites connected to fermion bags that change. If we randomly choose a spatial block containing about $30-60$ sites and focus on updating the bonds only within that block, during such a \textit{block-update} the size of the matrix $\Delta$ cannot be greater than the sum of the sites in the fermion bags that touch the sites within the block. We refer to this set of sites, which can be larger than the block size, as a \textit{super-bag} and denote its size as $s$. Since $\Delta$ is non-zero only in an $s\times s$ block, it is easy to show that the computation of $R$ (the ratio of the weight of the current configuration with that of the background configuration that existed at the time when the block update began) using (\ref{largem}), reduces to the computation of the determinant of an $s\times s$ matrix. Since $G_B$ and $M_T$ are fixed matrices during the entire \textit{block-update} they can be computed and stored and all proposals to update the current configuration within the block reduces to the computations of a determinant of an $s\times s$ matrix, independent of the system size \cite{Suppmat}.  

\begin{figure*}[t]
\minipage{0.32\textwidth}
  \includegraphics[width=\linewidth]{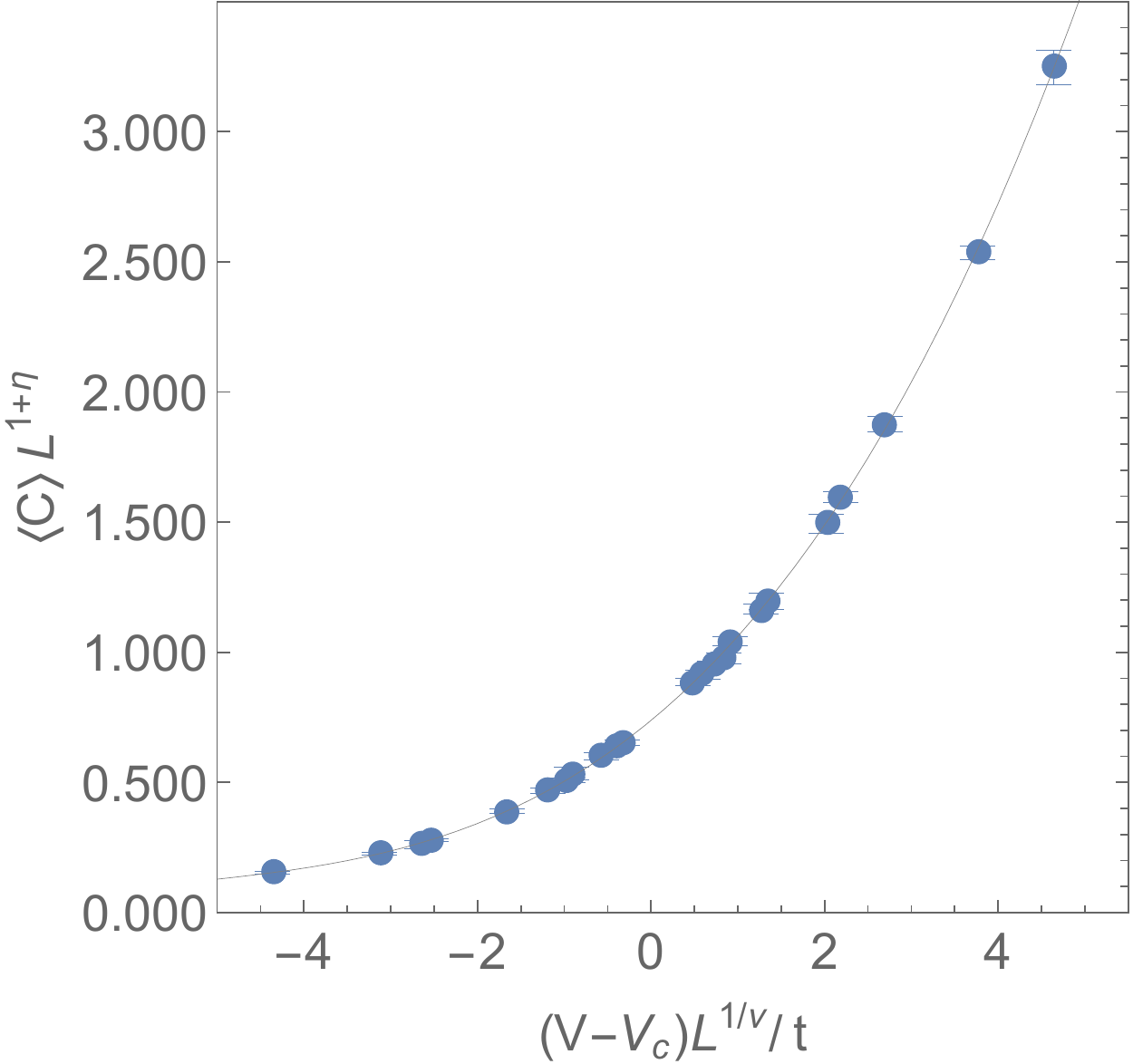}
  \caption{Critical scaling plot showing our Monte Carlo data scaled with $\eta=0.54$, $\nu=0.88$, $V_c=1.279 t$. The solid line shows $f(x)=0.77 + 0.30x+0.052x^2+0.0033x^3$. These values are obtained by a combined fit of the data in Table \ref{datatable} as explained in the text.}\label{fits}
\endminipage\hfill
\minipage{0.64\textwidth}
\hbox{
  \includegraphics[width=0.5\linewidth]{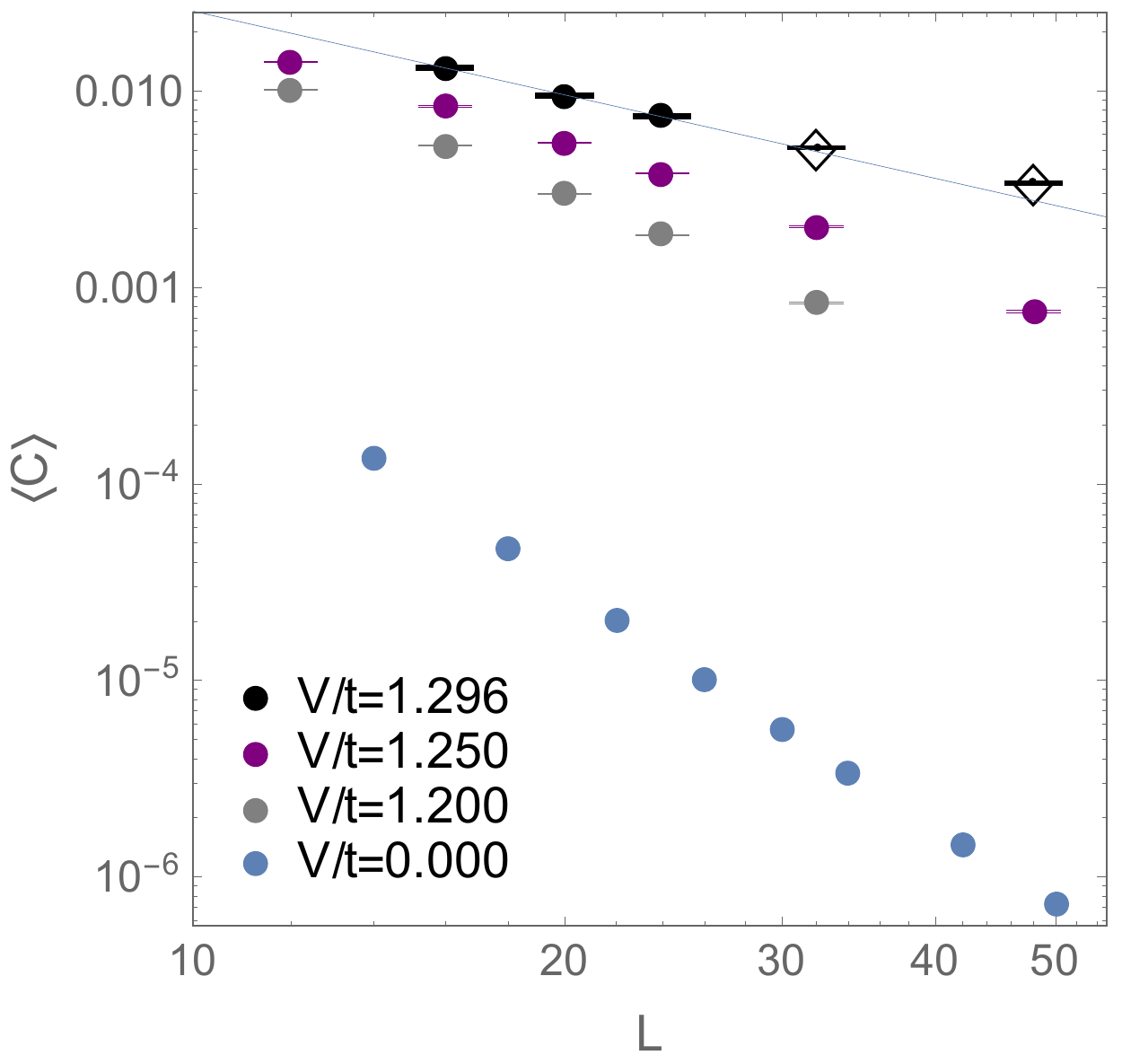}
  \includegraphics[width=0.5\linewidth]{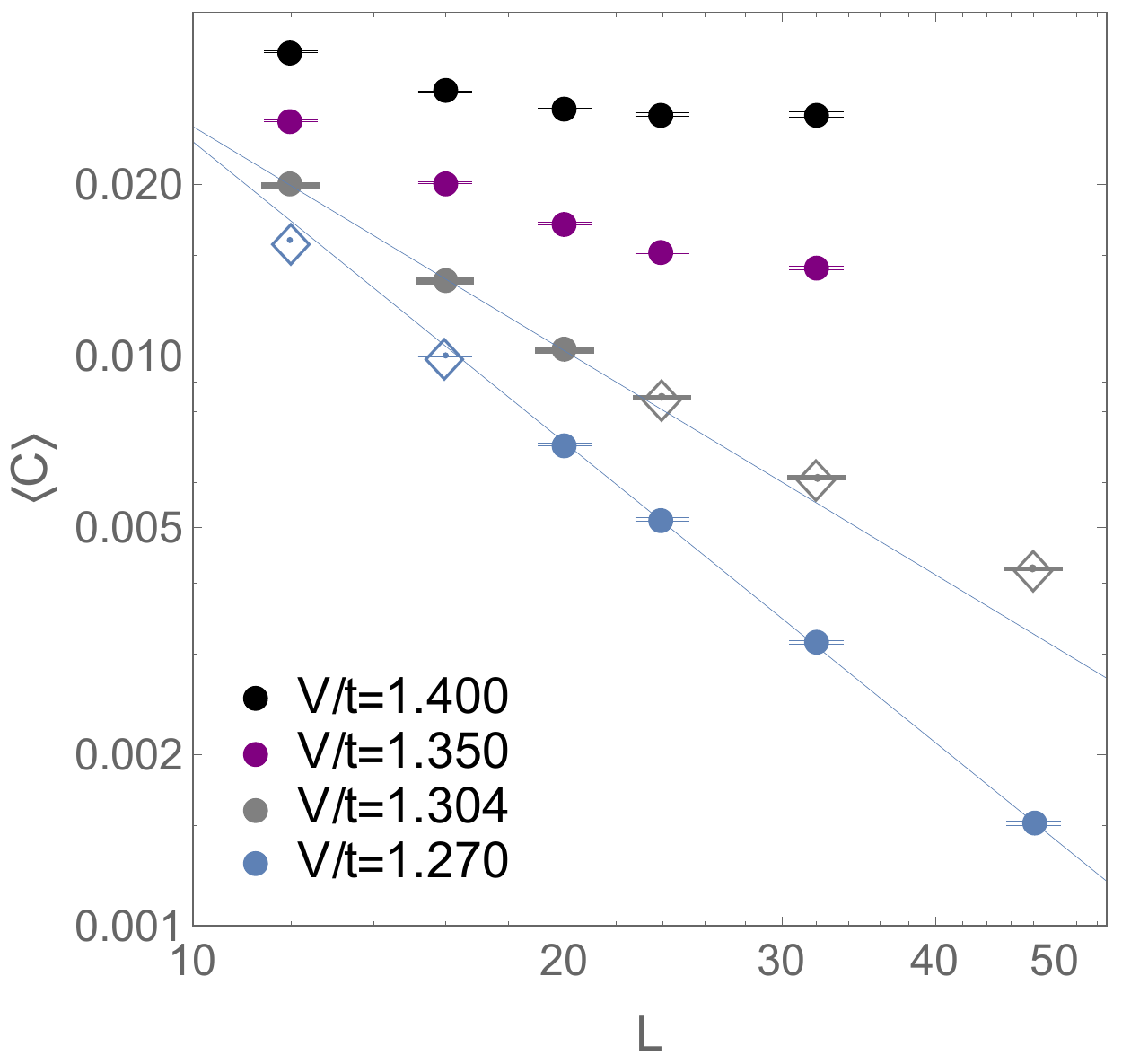}
  }
 \caption{Plots of $\langle C\rangle$ as a function of $L$ (with $\beta=L$) at various values of $V$. $\langle C\rangle$ scales as $L^{-4}$ at $V=0$ and saturates to a constant at $V=1.4 t$ as expected. At $V=V_c$ we  expect $\langle C\rangle \sim L^{-(1+\eta)}$. The solid lines show the best fits to this form at a fixed value of $V$ with open symbols being dropped from the fit. Previous predictions of $V_c$ were at $V=1.304 t$ \cite{1367-2630-16-10-103008} and  $V=1.296 t$ \cite{1367-2630-17-8-085003}. We find $\eta=0.31(2)$ and $\eta=0.41(4)$ at these couplings, consistent with previous results on small lattices. At $V=1.27t$ we find $\eta=0.74(2)$.}\label{vcscals}
\endminipage
\end{figure*}

Since the fermion bag size does not grow with system size the maximum size of $\Delta$ remains roughly the same on all time-slices even on large lattices. When the block within the same time-slice is changed, we need to recompute $G_B$ and $M_T$. Due to the structure of $G_B$ we can use identities such as 
\begin{eqnarray}
&& \left( \mathbbm{1}+M_1  M_2\right)^{-1} 
\ =\  (1-G_2) \nonumber \\
&& \quad \quad 
\Big((1-G_1)(1-G_2)+G_1G_2\Big)^{-1}(1-G_1)
\end{eqnarray}
where $G_i=(1+M_i)^{-1}M_i$, to express it in terms of partial $G_i$'s. These identities avoid instability issues. Since partial $G_i$'s can be calculated and stored we can compute $G_B$ easily without encountering instabilities. The recomputation of $G_B$ within a time-slice requires a time that scales as $O\left(sN^2\right)$ at most because we can use the Woodbury matrix identity in terms of inverses of stable partial products. When we change time-slices we use a storage scheme for our partial products similar to the one in \cite{Wang:2015rga} to facilitate updates that scale linearly in $\beta$. We have found that our algorithm not suffer from stabilization problems even when $N=10,000$ \cite{Suppmat}.

The time to complete a single sweep with our algorithm scales as $\beta N^3$, which is similar to the traditional auxiliary field algorithms. However, we believe we have reduced the prefactor significantly using the idea of fermion bags \cite{Suppmat}. In Fig.~\ref{sweepequil} we show equilibration of $N_b$ (the total number of bonds in a configuration) as a function of sweeps for $\beta=L=48,64,100$ and $V=1.304 t$. Although the $L=100$ data has not equilibrated, there is no bottleneck (see inset of Fig.~\ref{sweepequil}). We estimate the bond density at equilibrium to be $N_b/\beta L^2 \approx 2.7$, which means at $L=\beta=100$ we will have roughly $2.7$ million bonds after equilibration. A single sweep will then roughly require a month to complete on a single 3GHz CPU core. The results shown in the next section were obtained with an order of $10^4$ equilibrated configurations.

\section{Results at Criticality}

Using the algorithm described above, we have studied the two dimensional $t-V$ model and computed the critical exponents at the quantum phase transition between the massless and the massive fermion phases. These critical exponents are expected to belong to the Ising Gross-Neveu universality class with $N_f=1$ four-component Dirac fermions  \cite{PhysRevLett.86.958,Mihaila:2017ble}. For large values of $L$ we expect the observable $\langle C\rangle$ to scale as $L^{-4}$ in the massless phase and to saturate to a constant in the massive phase. In the critical region ($V\approx V_c$ and large values of $L$) we expect $\langle C\rangle$ to satisfy the leading critical finite size scaling relation \cite{Cardy:1988ag,PhysRevB.89.094516}
\begin{equation}
\langle C \rangle = \frac{1}{L^{1+\eta}} f\left(\left(V-V_c\right)L^{1/\nu}/t\right).
\end{equation}
Our Monte Carlo results are consistent with these expectations.

Table \ref{datatable} documents our results for $\langle C\rangle$ as a function of $V$ and $L$ near the critical point where we set $\beta=L$. Approximating $f(x)=f_0+f_1 x + f_2 x^2 + f_3 x^3$, we perform a seven parameter combined fit of all the data given in Table \ref{datatable}, except the $L=32$ data at $V=1.4$. This point does not seem to lie within the scaling window. Using the fit we obtain $\eta=0.54(6)$, $\nu=0.88(2)$, $V_c=1.279(3) t$, $f_0=0.77(11)$, $f_1=0.30(4)$, $f_2=0.052(8)$ and $f_4=0.0033(6)$. The $\chi^2/DOF$ for the fit is $0.8$. We show the data and the scaling fit in the left plot of Fig.~\ref{fits}. Theoretical predictions for the exponents are compatible with our results \cite{PhysRevLett.86.958,Mihaila:2017ble}.

\begin{table}[h]
\begin{tabular}[t]{|l||l|l|l|l|}
\hline
 $V/t$ & $L=20$ & $L=24$ & $L=32$ & $L=48$ \\
\hline
$1.200$ & 0.00298(3) & 0.00184(3) &
0.00080(1) & $\qquad -$ \\
\hline
$1.250$ & 0.00545(6) & 0.00380(5) & 0.00204(2) & 0.00074(2) \\
\hline
$1.270$ & 0.00699(8) & 0.00517(7) & 0.00315(4) & 0.00151(3)  \\
\hline
$1.296$ & 0.00946(10) & 0.00740(9) 
& 0.00512(6) & 0.00339(5)  \\
\hline
$1.304$ & 0.01022(8) & 0.00844(9) 
& 0.00611(6) & 0.00423(5)  \\
\hline
$1.350$ & 0.01705(16) & 0.01522(16) 
& 0.01426(18) & $\qquad -$\\
\hline
$1.400$ & 0.02707(20) & 0.02630(35) & 0.02637(38) & $\qquad -$\\
\hline
\end{tabular}
\caption{Our Monte Carlo results for the $t$-$V$ model (\ref{tvmodel}) on a square lattice with $20\leq L \leq 48$ and $\beta=L$.}
\label{datatable}
\end{table}

The $t-V$ model we study here has been studied earlier on smaller lattices by two groups. Not surprisingly, the critical point and the critical exponents measured are in disagreement with each other. The first calculation was performed on lattices up to $N=400$ sites and it was found that $V_c=1.304(2)$, $\eta=0.318(8)$ and $\nu=0.80(6)$ \cite{1367-2630-16-10-103008}. In a later calculation lattices up to $N=484$ sites were used and it was found that $V_c=1.296(1)$, $\eta=0.43(2)$ and $\nu=0.79(4)$ \cite{1367-2630-17-8-085003}.

Our results are obtained from lattice sizes that are five times larger than earlier studies and suggest a lower critical point and consequently a higher value for the exponent $\eta$. The value of $\nu$ also seems slightly higher but not inconsistent with previous results. If we exclude the larger lattice results we do find consistency with previous results. For example, if we assume $V_c/t=1.296$ or $1.304$ and fit our data to the form $L^{-(1+\eta)}$, after dropping larger values of $L$
we get $\eta=0.41(4)$ and $\eta=0.31(4)$ respectively with a reasonable $\chi^2/DOF$ (see Fig.~\ref{vcscals}). Unfortunately, the fits fail dramatically if $L=32$ and $L=48$. On the other hand at $V = 1.27 t$ the data fits well if we drop smaller values of $L$ and gives us $\eta=0.74(2)$. If we force $V_c=1.27 t$ in the combined fit, the $\chi^2/DOF$ increases to $1.3$.

\section{Conclusions}

In this work we have demonstrated that the idea of fermion bags can be combined with standard Monte Carlo techniques to study large system sizes in continuous time. We studied the quantum critical behavior in the simplest Ising Gross-Neveu universality class and extracted the critical exponents using lattice sizes that were five times larger than previous work. Even larger sizes are feasible with supercomputers available today. With additional research, the idea of fermion bags should be applicable to a wide class of models.

\section*{Acknowledgments}

We thank F.~Assaad, K.~Damle, T.~Grover, L.~Wang, S.~Wessel and U.-J.~Wiese for stimulating discussions. The work is supported by the U.S. Department of Energy, Office of Science, Nuclear Physics program under Award Number DE-FG02-05ER41368. Calculations on large lattices were made possible by the Open Science Grid, which is supported by the National Science Foundation.

\bibliographystyle{apsrev4-1}
\bibliography{refs}

\begin{thebibliography}{49}%
\makeatletter
\providecommand \@ifxundefined [1]{%
 \@ifx{#1\undefined}
}%
\providecommand \@ifnum [1]{%
 \ifnum #1\expandafter \@firstoftwo
 \else \expandafter \@secondoftwo
 \fi
}%
\providecommand \@ifx [1]{%
 \ifx #1\expandafter \@firstoftwo
 \else \expandafter \@secondoftwo
 \fi
}%
\providecommand \natexlab [1]{#1}%
\providecommand \enquote  [1]{``#1''}%
\providecommand \bibnamefont  [1]{#1}%
\providecommand \bibfnamefont [1]{#1}%
\providecommand \citenamefont [1]{#1}%
\providecommand \href@noop [0]{\@secondoftwo}%
\providecommand \href [0]{\begingroup \@sanitize@url \@href}%
\providecommand \@href[1]{\@@startlink{#1}\@@href}%
\providecommand \@@href[1]{\endgroup#1\@@endlink}%
\providecommand \@sanitize@url [0]{\catcode `\\12\catcode `\$12\catcode
  `\&12\catcode `\#12\catcode `\^12\catcode `\_12\catcode `\%12\relax}%
\providecommand \@@startlink[1]{}%
\providecommand \@@endlink[0]{}%
\providecommand \url  [0]{\begingroup\@sanitize@url \@url }%
\providecommand \@url [1]{\endgroup\@href {#1}{\urlprefix }}%
\providecommand \urlprefix  [0]{URL }%
\providecommand \Eprint [0]{\href }%
\providecommand \doibase [0]{http://dx.doi.org/}%
\providecommand \selectlanguage [0]{\@gobble}%
\providecommand \bibinfo  [0]{\@secondoftwo}%
\providecommand \bibfield  [0]{\@secondoftwo}%
\providecommand \translation [1]{[#1]}%
\providecommand \BibitemOpen [0]{}%
\providecommand \bibitemStop [0]{}%
\providecommand \bibitemNoStop [0]{.\EOS\space}%
\providecommand \EOS [0]{\spacefactor3000\relax}%
\providecommand \BibitemShut  [1]{\csname bibitem#1\endcsname}%
\let\auto@bib@innerbib\@empty
%</preamble>
\bibitem [{\citenamefont {Troyer}\ and\ \citenamefont
  {Wiese}(2005)}]{Troyer:2004ge}%
  \BibitemOpen
  \bibfield  {author} {\bibinfo {author} {\bibfnamefont {M.}~\bibnamefont
  {Troyer}}\ and\ \bibinfo {author} {\bibfnamefont {U.-J.}\ \bibnamefont
  {Wiese}},\ }\href {\doibase 10.1103/PhysRevLett.94.170201} {\bibfield
  {journal} {\bibinfo  {journal} {Phys. Rev. Lett.}\ }\textbf {\bibinfo
  {volume} {94}},\ \bibinfo {pages} {170201} (\bibinfo {year} {2005})},\
  \Eprint {http://arxiv.org/abs/cond-mat/0408370} {arXiv:cond-mat/0408370
  [cond-mat]} \BibitemShut {NoStop}%
%%CITATION = COND-MAT/0408370;%%
\bibitem [{\citenamefont {Rosenstein}\ \emph {et~al.}(1989)\citenamefont
  {Rosenstein}, \citenamefont {Warr},\ and\ \citenamefont
  {Park}}]{PhysRevLett.62.1433}%
  \BibitemOpen
  \bibfield  {author} {\bibinfo {author} {\bibfnamefont {B.}~\bibnamefont
  {Rosenstein}}, \bibinfo {author} {\bibfnamefont {B.~J.}\ \bibnamefont
  {Warr}}, \ and\ \bibinfo {author} {\bibfnamefont {S.~H.}\ \bibnamefont
  {Park}},\ }\href {\doibase 10.1103/PhysRevLett.62.1433} {\bibfield  {journal}
  {\bibinfo  {journal} {Phys. Rev. Lett.}\ }\textbf {\bibinfo {volume} {62}},\
  \bibinfo {pages} {1433} (\bibinfo {year} {1989})}\BibitemShut {NoStop}%
\bibitem [{\citenamefont {Rosenstein}\ \emph {et~al.}(1991)\citenamefont
  {Rosenstein}, \citenamefont {Warr},\ and\ \citenamefont
  {Park}}]{Rosenstein199159}%
  \BibitemOpen
  \bibfield  {author} {\bibinfo {author} {\bibfnamefont {B.}~\bibnamefont
  {Rosenstein}}, \bibinfo {author} {\bibfnamefont {B.~J.}\ \bibnamefont
  {Warr}}, \ and\ \bibinfo {author} {\bibfnamefont {S.~H.}\ \bibnamefont
  {Park}},\ }\href {\doibase http://dx.doi.org/10.1016/0370-1573(91)90129-A}
  {\bibfield  {journal} {\bibinfo  {journal} {Physics Reports}\ }\textbf
  {\bibinfo {volume} {205}},\ \bibinfo {pages} {59 } (\bibinfo {year}
  {1991})}\BibitemShut {NoStop}%
\bibitem [{\citenamefont {Herbut}(2006)}]{PhysRevLett.97.146401}%
  \BibitemOpen
  \bibfield  {author} {\bibinfo {author} {\bibfnamefont {I.~F.}\ \bibnamefont
  {Herbut}},\ }\href {\doibase 10.1103/PhysRevLett.97.146401} {\bibfield
  {journal} {\bibinfo  {journal} {Phys. Rev. Lett.}\ }\textbf {\bibinfo
  {volume} {97}},\ \bibinfo {pages} {146401} (\bibinfo {year}
  {2006})}\BibitemShut {NoStop}%
\bibitem [{\citenamefont {Janssen}\ and\ \citenamefont
  {Gies}(2012)}]{PhysRevD.86.105007}%
  \BibitemOpen
  \bibfield  {author} {\bibinfo {author} {\bibfnamefont {L.}~\bibnamefont
  {Janssen}}\ and\ \bibinfo {author} {\bibfnamefont {H.}~\bibnamefont {Gies}},\
  }\href {\doibase 10.1103/PhysRevD.86.105007} {\bibfield  {journal} {\bibinfo
  {journal} {Phys. Rev. D}\ }\textbf {\bibinfo {volume} {86}},\ \bibinfo
  {pages} {105007} (\bibinfo {year} {2012})}\BibitemShut {NoStop}%
\bibitem [{\citenamefont {Wellegehausen}\ \emph {et~al.}(2017)\citenamefont
  {Wellegehausen}, \citenamefont {Schmidt},\ and\ \citenamefont
  {Wipf}}]{Wellegehausen:2017goy}%
  \BibitemOpen
  \bibfield  {author} {\bibinfo {author} {\bibfnamefont {B.~H.}\ \bibnamefont
  {Wellegehausen}}, \bibinfo {author} {\bibfnamefont {D.}~\bibnamefont
  {Schmidt}}, \ and\ \bibinfo {author} {\bibfnamefont {A.}~\bibnamefont
  {Wipf}},\ }\href@noop {} {\  (\bibinfo {year} {2017})},\ \Eprint
  {http://arxiv.org/abs/1708.01160} {arXiv:1708.01160 [hep-lat]} \BibitemShut
  {NoStop}%
%%CITATION = ARXIV:1708.01160;%%
\bibitem [{\citenamefont {You}\ \emph {et~al.}(2017)\citenamefont {You},
  \citenamefont {He}, \citenamefont {Xu},\ and\ \citenamefont
  {Vishwanath}}]{You:2017ltx}%
  \BibitemOpen
  \bibfield  {author} {\bibinfo {author} {\bibfnamefont {Y.-Z.}\ \bibnamefont
  {You}}, \bibinfo {author} {\bibfnamefont {Y.-C.}\ \bibnamefont {He}},
  \bibinfo {author} {\bibfnamefont {C.}~\bibnamefont {Xu}}, \ and\ \bibinfo
  {author} {\bibfnamefont {A.}~\bibnamefont {Vishwanath}},\ }\href@noop {} {\
  (\bibinfo {year} {2017})},\ \Eprint {http://arxiv.org/abs/1705.09313}
  {arXiv:1705.09313 [cond-mat.str-el]} \BibitemShut {NoStop}%
%%CITATION = ARXIV:1705.09313;%%
\bibitem [{\citenamefont {Hands}\ \emph {et~al.}(1993)\citenamefont {Hands},
  \citenamefont {Kocic},\ and\ \citenamefont {Kogut}}]{Hands199329}%
  \BibitemOpen
  \bibfield  {author} {\bibinfo {author} {\bibfnamefont {S.}~\bibnamefont
  {Hands}}, \bibinfo {author} {\bibfnamefont {A.}~\bibnamefont {Kocic}}, \ and\
  \bibinfo {author} {\bibfnamefont {J.}~\bibnamefont {Kogut}},\ }\href
  {\doibase http://dx.doi.org/10.1006/aphy.1993.1039} {\bibfield  {journal}
  {\bibinfo  {journal} {Annals of Physics}\ }\textbf {\bibinfo {volume}
  {224}},\ \bibinfo {pages} {29 } (\bibinfo {year} {1993})}\BibitemShut
  {NoStop}%
\bibitem [{\citenamefont {Kärkkäinen}\ \emph {et~al.}(1994)\citenamefont
  {Kärkkäinen}, \citenamefont {Lacaze}, \citenamefont {Lacock},\ and\
  \citenamefont {Petersson}}]{Karkkainen94}%
  \BibitemOpen
  \bibfield  {author} {\bibinfo {author} {\bibfnamefont {L.}~\bibnamefont
  {Kärkkäinen}}, \bibinfo {author} {\bibfnamefont {R.}~\bibnamefont
  {Lacaze}}, \bibinfo {author} {\bibfnamefont {P.}~\bibnamefont {Lacock}}, \
  and\ \bibinfo {author} {\bibfnamefont {B.}~\bibnamefont {Petersson}},\ }\href
  {\doibase http://dx.doi.org/10.1016/0550-3213(94)90309-3} {\bibfield
  {journal} {\bibinfo  {journal} {Nuclear Physics B}\ }\textbf {\bibinfo
  {volume} {415}},\ \bibinfo {pages} {781 } (\bibinfo {year}
  {1994})}\BibitemShut {NoStop}%
\bibitem [{\citenamefont {Focht}\ \emph {et~al.}(1996)\citenamefont {Focht},
  \citenamefont {Jers\'ak},\ and\ \citenamefont {Paul}}]{PhysRevD.53.4616}%
  \BibitemOpen
  \bibfield  {author} {\bibinfo {author} {\bibfnamefont {E.}~\bibnamefont
  {Focht}}, \bibinfo {author} {\bibfnamefont {J.}~\bibnamefont {Jers\'ak}}, \
  and\ \bibinfo {author} {\bibfnamefont {J.}~\bibnamefont {Paul}},\ }\href
  {\doibase 10.1103/PhysRevD.53.4616} {\bibfield  {journal} {\bibinfo
  {journal} {Phys. Rev. D}\ }\textbf {\bibinfo {volume} {53}},\ \bibinfo
  {pages} {4616} (\bibinfo {year} {1996})}\BibitemShut {NoStop}%
\bibitem [{\citenamefont {Christofi}\ and\ \citenamefont
  {Strouthos}(2007)}]{Stavros07}%
  \BibitemOpen
  \bibfield  {author} {\bibinfo {author} {\bibfnamefont {S.}~\bibnamefont
  {Christofi}}\ and\ \bibinfo {author} {\bibfnamefont {C.}~\bibnamefont
  {Strouthos}},\ }\href {http://stacks.iop.org/1126-6708/2007/i=05/a=088}
  {\bibfield  {journal} {\bibinfo  {journal} {Journal of High Energy Physics}\
  }\textbf {\bibinfo {volume} {2007}},\ \bibinfo {pages} {088} (\bibinfo {year}
  {2007})}\BibitemShut {NoStop}%
\bibitem [{\citenamefont {Drut}\ and\ \citenamefont
  {L\"ahde}(2009)}]{PhysRevB.79.165425}%
  \BibitemOpen
  \bibfield  {author} {\bibinfo {author} {\bibfnamefont {J.~E.}\ \bibnamefont
  {Drut}}\ and\ \bibinfo {author} {\bibfnamefont {T.~A.}\ \bibnamefont
  {L\"ahde}},\ }\href {\doibase 10.1103/PhysRevB.79.165425} {\bibfield
  {journal} {\bibinfo  {journal} {Phys. Rev. B}\ }\textbf {\bibinfo {volume}
  {79}},\ \bibinfo {pages} {165425} (\bibinfo {year} {2009})}\BibitemShut
  {NoStop}%
\bibitem [{\citenamefont {Karthik}\ and\ \citenamefont
  {Narayanan}(2016)}]{PhysRevD.94.065026}%
  \BibitemOpen
  \bibfield  {author} {\bibinfo {author} {\bibfnamefont {N.}~\bibnamefont
  {Karthik}}\ and\ \bibinfo {author} {\bibfnamefont {R.}~\bibnamefont
  {Narayanan}},\ }\href {\doibase 10.1103/PhysRevD.94.065026} {\bibfield
  {journal} {\bibinfo  {journal} {Phys. Rev. D}\ }\textbf {\bibinfo {volume}
  {94}},\ \bibinfo {pages} {065026} (\bibinfo {year} {2016})}\BibitemShut
  {NoStop}%
\bibitem [{\citenamefont {Hands}(2016)}]{Hands:2016foa}%
  \BibitemOpen
  \bibfield  {author} {\bibinfo {author} {\bibfnamefont {S.}~\bibnamefont
  {Hands}},\ }\href {\doibase 10.1007/JHEP11(2016)015} {\bibfield  {journal}
  {\bibinfo  {journal} {JHEP}\ }\textbf {\bibinfo {volume} {11}},\ \bibinfo
  {pages} {015} (\bibinfo {year} {2016})},\ \Eprint
  {http://arxiv.org/abs/1610.04394} {arXiv:1610.04394 [hep-lat]} \BibitemShut
  {NoStop}%
%%CITATION = ARXIV:1610.04394;%%
\bibitem [{\citenamefont {Hands}(2017)}]{Hands:2017hhk}%
  \BibitemOpen
  \bibfield  {author} {\bibinfo {author} {\bibfnamefont {S.}~\bibnamefont
  {Hands}},\ }in\ \href
  {https://inspirehep.net/record/1620058/files/arXiv:1708.07686.pdf} {\emph
  {\bibinfo {booktitle} {{35th International Symposium on Lattice Field Theory
  (Lattice 2017) Granada, Spain, June 18-24, 2017}}}}\ (\bibinfo {year}
  {2017})\ \Eprint {http://arxiv.org/abs/1708.07686} {arXiv:1708.07686
  [hep-lat]} \BibitemShut {NoStop}%
%%CITATION = ARXIV:1708.07686;%%
\bibitem [{\citenamefont {Rubtsov}\ \emph {et~al.}(2005)\citenamefont
  {Rubtsov}, \citenamefont {Savkin},\ and\ \citenamefont
  {Lichtenstein}}]{PhysRevB.72.035122}%
  \BibitemOpen
  \bibfield  {author} {\bibinfo {author} {\bibfnamefont {A.~N.}\ \bibnamefont
  {Rubtsov}}, \bibinfo {author} {\bibfnamefont {V.~V.}\ \bibnamefont {Savkin}},
  \ and\ \bibinfo {author} {\bibfnamefont {A.~I.}\ \bibnamefont
  {Lichtenstein}},\ }\href {\doibase 10.1103/PhysRevB.72.035122} {\bibfield
  {journal} {\bibinfo  {journal} {Phys. Rev. B}\ }\textbf {\bibinfo {volume}
  {72}},\ \bibinfo {pages} {035122} (\bibinfo {year} {2005})}\BibitemShut
  {NoStop}%
\bibitem [{\citenamefont {Gull}\ \emph {et~al.}(2011)\citenamefont {Gull},
  \citenamefont {Millis}, \citenamefont {Lichtenstein}, \citenamefont
  {Rubtsov}, \citenamefont {Troyer},\ and\ \citenamefont
  {Werner}}]{RevModPhys.83.349}%
  \BibitemOpen
  \bibfield  {author} {\bibinfo {author} {\bibfnamefont {E.}~\bibnamefont
  {Gull}}, \bibinfo {author} {\bibfnamefont {A.~J.}\ \bibnamefont {Millis}},
  \bibinfo {author} {\bibfnamefont {A.~I.}\ \bibnamefont {Lichtenstein}},
  \bibinfo {author} {\bibfnamefont {A.~N.}\ \bibnamefont {Rubtsov}}, \bibinfo
  {author} {\bibfnamefont {M.}~\bibnamefont {Troyer}}, \ and\ \bibinfo {author}
  {\bibfnamefont {P.}~\bibnamefont {Werner}},\ }\href {\doibase
  10.1103/RevModPhys.83.349} {\bibfield  {journal} {\bibinfo  {journal} {Rev.
  Mod. Phys.}\ }\textbf {\bibinfo {volume} {83}},\ \bibinfo {pages} {349}
  (\bibinfo {year} {2011})}\BibitemShut {NoStop}%
\bibitem [{\citenamefont {Wang}\ \emph {et~al.}(2015)\citenamefont {Wang},
  \citenamefont {Iazzi}, \citenamefont {Corboz},\ and\ \citenamefont
  {Troyer}}]{Wang:2015rga}%
  \BibitemOpen
  \bibfield  {author} {\bibinfo {author} {\bibfnamefont {L.}~\bibnamefont
  {Wang}}, \bibinfo {author} {\bibfnamefont {M.}~\bibnamefont {Iazzi}},
  \bibinfo {author} {\bibfnamefont {P.}~\bibnamefont {Corboz}}, \ and\ \bibinfo
  {author} {\bibfnamefont {M.}~\bibnamefont {Troyer}},\ }\href {\doibase
  10.1103/PhysRevB.91.235151} {\bibfield  {journal} {\bibinfo  {journal} {Phys.
  Rev.}\ }\textbf {\bibinfo {volume} {B91}},\ \bibinfo {pages} {235151}
  (\bibinfo {year} {2015})},\ \Eprint {http://arxiv.org/abs/1501.00986}
  {arXiv:1501.00986 [cond-mat.str-el]} \BibitemShut {NoStop}%
%%CITATION = ARXIV:1501.00986;%%
\bibitem [{\citenamefont {Blankenbecler}\ \emph {et~al.}(1981)\citenamefont
  {Blankenbecler}, \citenamefont {Scalapino},\ and\ \citenamefont
  {Sugar}}]{PhysRevD.24.2278}%
  \BibitemOpen
  \bibfield  {author} {\bibinfo {author} {\bibfnamefont {R.}~\bibnamefont
  {Blankenbecler}}, \bibinfo {author} {\bibfnamefont {D.~J.}\ \bibnamefont
  {Scalapino}}, \ and\ \bibinfo {author} {\bibfnamefont {R.~L.}\ \bibnamefont
  {Sugar}},\ }\href {\doibase 10.1103/PhysRevD.24.2278} {\bibfield  {journal}
  {\bibinfo  {journal} {Phys. Rev. D}\ }\textbf {\bibinfo {volume} {24}},\
  \bibinfo {pages} {2278} (\bibinfo {year} {1981})}\BibitemShut {NoStop}%
\bibitem [{\citenamefont {Bercx}\ \emph {et~al.}(2017)\citenamefont {Bercx},
  \citenamefont {Goth}, \citenamefont {Hofmann},\ and\ \citenamefont
  {Assaad}}]{Bercx:2017pit}%
  \BibitemOpen
  \bibfield  {author} {\bibinfo {author} {\bibfnamefont {M.}~\bibnamefont
  {Bercx}}, \bibinfo {author} {\bibfnamefont {F.}~\bibnamefont {Goth}},
  \bibinfo {author} {\bibfnamefont {J.~S.}\ \bibnamefont {Hofmann}}, \ and\
  \bibinfo {author} {\bibfnamefont {F.~F.}\ \bibnamefont {Assaad}},\
  }\href@noop {} {\  (\bibinfo {year} {2017})},\ \Eprint
  {http://arxiv.org/abs/1704.00131} {arXiv:1704.00131 [cond-mat.str-el]}
  \BibitemShut {NoStop}%
\bibitem [{\citenamefont {Wang}\ \emph {et~al.}(2014)\citenamefont {Wang},
  \citenamefont {Corboz},\ and\ \citenamefont
  {Troyer}}]{1367-2630-16-10-103008}%
  \BibitemOpen
  \bibfield  {author} {\bibinfo {author} {\bibfnamefont {L.}~\bibnamefont
  {Wang}}, \bibinfo {author} {\bibfnamefont {P.}~\bibnamefont {Corboz}}, \ and\
  \bibinfo {author} {\bibfnamefont {M.}~\bibnamefont {Troyer}},\ }\href
  {http://stacks.iop.org/1367-2630/16/i=10/a=103008} {\bibfield  {journal}
  {\bibinfo  {journal} {New Journal of Physics}\ }\textbf {\bibinfo {volume}
  {16}},\ \bibinfo {pages} {103008} (\bibinfo {year} {2014})}\BibitemShut
  {NoStop}%
\bibitem [{\citenamefont {Li}\ \emph {et~al.}(2015{\natexlab{a}})\citenamefont
  {Li}, \citenamefont {Jiang},\ and\ \citenamefont
  {Yao}}]{1367-2630-17-8-085003}%
  \BibitemOpen
  \bibfield  {author} {\bibinfo {author} {\bibfnamefont {Z.-X.}\ \bibnamefont
  {Li}}, \bibinfo {author} {\bibfnamefont {Y.-F.}\ \bibnamefont {Jiang}}, \
  and\ \bibinfo {author} {\bibfnamefont {H.}~\bibnamefont {Yao}},\ }\href
  {http://stacks.iop.org/1367-2630/17/i=8/a=085003} {\bibfield  {journal}
  {\bibinfo  {journal} {New Journal of Physics}\ }\textbf {\bibinfo {volume}
  {17}},\ \bibinfo {pages} {085003} (\bibinfo {year}
  {2015}{\natexlab{a}})}\BibitemShut {NoStop}%
\bibitem [{\citenamefont {Hesselmann}\ and\ \citenamefont
  {Wessel}(2016)}]{Hesselmann:2016tvh}%
  \BibitemOpen
  \bibfield  {author} {\bibinfo {author} {\bibfnamefont {S.}~\bibnamefont
  {Hesselmann}}\ and\ \bibinfo {author} {\bibfnamefont {S.}~\bibnamefont
  {Wessel}},\ }\href {\doibase 10.1103/PhysRevB.93.155157} {\bibfield
  {journal} {\bibinfo  {journal} {Phys. Rev.}\ }\textbf {\bibinfo {volume}
  {B93}},\ \bibinfo {pages} {155157} (\bibinfo {year} {2016})},\ \Eprint
  {http://arxiv.org/abs/1602.02096} {arXiv:1602.02096 [cond-mat.str-el]}
  \BibitemShut {NoStop}%
%%CITATION = ARXIV:1602.02096;%%
\bibitem [{\citenamefont {Otsuka}\ \emph {et~al.}(2016)\citenamefont {Otsuka},
  \citenamefont {Yunoki},\ and\ \citenamefont {Sorella}}]{PhysRevX.6.011029}%
  \BibitemOpen
  \bibfield  {author} {\bibinfo {author} {\bibfnamefont {Y.}~\bibnamefont
  {Otsuka}}, \bibinfo {author} {\bibfnamefont {S.}~\bibnamefont {Yunoki}}, \
  and\ \bibinfo {author} {\bibfnamefont {S.}~\bibnamefont {Sorella}},\ }\href
  {\doibase 10.1103/PhysRevX.6.011029} {\bibfield  {journal} {\bibinfo
  {journal} {Phys. Rev. X}\ }\textbf {\bibinfo {volume} {6}},\ \bibinfo {pages}
  {011029} (\bibinfo {year} {2016})}\BibitemShut {NoStop}%
\bibitem [{\citenamefont {Ulybyshev}\ \emph {et~al.}(2013)\citenamefont
  {Ulybyshev}, \citenamefont {Buividovich}, \citenamefont {Katsnelson},\ and\
  \citenamefont {Polikarpov}}]{PhysRevLett.111.056801}%
  \BibitemOpen
  \bibfield  {author} {\bibinfo {author} {\bibfnamefont {M.~V.}\ \bibnamefont
  {Ulybyshev}}, \bibinfo {author} {\bibfnamefont {P.~V.}\ \bibnamefont
  {Buividovich}}, \bibinfo {author} {\bibfnamefont {M.~I.}\ \bibnamefont
  {Katsnelson}}, \ and\ \bibinfo {author} {\bibfnamefont {M.~I.}\ \bibnamefont
  {Polikarpov}},\ }\href {\doibase 10.1103/PhysRevLett.111.056801} {\bibfield
  {journal} {\bibinfo  {journal} {Phys. Rev. Lett.}\ }\textbf {\bibinfo
  {volume} {111}},\ \bibinfo {pages} {056801} (\bibinfo {year}
  {2013})}\BibitemShut {NoStop}%
\bibitem [{\citenamefont {Körner}\ \emph {et~al.}(2017)\citenamefont
  {Körner}, \citenamefont {Smith}, \citenamefont {Buividovich}, \citenamefont
  {Ulybyshev},\ and\ \citenamefont {von Smekal}}]{Korner:2017qhf}%
  \BibitemOpen
  \bibfield  {author} {\bibinfo {author} {\bibfnamefont {M.}~\bibnamefont
  {Körner}}, \bibinfo {author} {\bibfnamefont {D.}~\bibnamefont {Smith}},
  \bibinfo {author} {\bibfnamefont {P.}~\bibnamefont {Buividovich}}, \bibinfo
  {author} {\bibfnamefont {M.}~\bibnamefont {Ulybyshev}}, \ and\ \bibinfo
  {author} {\bibfnamefont {L.}~\bibnamefont {von Smekal}},\ }\href@noop {} {\
  (\bibinfo {year} {2017})},\ \Eprint {http://arxiv.org/abs/1704.03757}
  {arXiv:1704.03757 [cond-mat.str-el]} \BibitemShut {NoStop}%
%%CITATION = ARXIV:1704.03757;%%
\bibitem [{\citenamefont {Beyl}\ \emph {et~al.}(2017)\citenamefont {Beyl},
  \citenamefont {Goth},\ and\ \citenamefont {Assaad}}]{Beyl:2017kwp}%
  \BibitemOpen
  \bibfield  {author} {\bibinfo {author} {\bibfnamefont {S.}~\bibnamefont
  {Beyl}}, \bibinfo {author} {\bibfnamefont {F.}~\bibnamefont {Goth}}, \ and\
  \bibinfo {author} {\bibfnamefont {F.~F.}\ \bibnamefont {Assaad}},\
  }\href@noop {} {\  (\bibinfo {year} {2017})},\ \Eprint
  {http://arxiv.org/abs/1708.03661} {arXiv:1708.03661 [cond-mat.str-el]}
  \BibitemShut {NoStop}%
%%CITATION = ARXIV:1708.03661;%%
\bibitem [{\citenamefont {Chandrasekharan}(2010)}]{PhysRevD.82.025007}%
  \BibitemOpen
  \bibfield  {author} {\bibinfo {author} {\bibfnamefont {S.}~\bibnamefont
  {Chandrasekharan}},\ }\href {\doibase 10.1103/PhysRevD.82.025007} {\bibfield
  {journal} {\bibinfo  {journal} {Phys. Rev. D}\ }\textbf {\bibinfo {volume}
  {82}},\ \bibinfo {pages} {025007} (\bibinfo {year} {2010})}\BibitemShut
  {NoStop}%
\bibitem [{\citenamefont {Chandrasekharan}(2013)}]{Chandrasekharan2013}%
  \BibitemOpen
  \bibfield  {author} {\bibinfo {author} {\bibfnamefont {S.}~\bibnamefont
  {Chandrasekharan}},\ }\href {\doibase 10.1140/epja/i2013-13090-y} {\bibfield
  {journal} {\bibinfo  {journal} {The European Physical Journal A}\ }\textbf
  {\bibinfo {volume} {49}},\ \bibinfo {pages} {90} (\bibinfo {year}
  {2013})}\BibitemShut {NoStop}%
\bibitem [{\citenamefont {Chandrasekharan}\ and\ \citenamefont
  {Li}(2012)}]{PhysRevLett.108.140404}%
  \BibitemOpen
  \bibfield  {author} {\bibinfo {author} {\bibfnamefont {S.}~\bibnamefont
  {Chandrasekharan}}\ and\ \bibinfo {author} {\bibfnamefont {A.}~\bibnamefont
  {Li}},\ }\href {\doibase 10.1103/PhysRevLett.108.140404} {\bibfield
  {journal} {\bibinfo  {journal} {Phys. Rev. Lett.}\ }\textbf {\bibinfo
  {volume} {108}},\ \bibinfo {pages} {140404} (\bibinfo {year}
  {2012})}\BibitemShut {NoStop}%
\bibitem [{\citenamefont {Chandrasekharan}\ and\ \citenamefont
  {Li}(2013)}]{PhysRevD.88.021701}%
  \BibitemOpen
  \bibfield  {author} {\bibinfo {author} {\bibfnamefont {S.}~\bibnamefont
  {Chandrasekharan}}\ and\ \bibinfo {author} {\bibfnamefont {A.}~\bibnamefont
  {Li}},\ }\href {\doibase 10.1103/PhysRevD.88.021701} {\bibfield  {journal}
  {\bibinfo  {journal} {Phys. Rev. D}\ }\textbf {\bibinfo {volume} {88}},\
  \bibinfo {pages} {021701} (\bibinfo {year} {2013})}\BibitemShut {NoStop}%
\bibitem [{\citenamefont {Chandrasekharan}\ and\ \citenamefont
  {Wiese}(1999)}]{PhysRevLett.83.3116}%
  \BibitemOpen
  \bibfield  {author} {\bibinfo {author} {\bibfnamefont {S.}~\bibnamefont
  {Chandrasekharan}}\ and\ \bibinfo {author} {\bibfnamefont {U.-J.}\
  \bibnamefont {Wiese}},\ }\href {\doibase 10.1103/PhysRevLett.83.3116}
  {\bibfield  {journal} {\bibinfo  {journal} {Phys. Rev. Lett.}\ }\textbf
  {\bibinfo {volume} {83}},\ \bibinfo {pages} {3116} (\bibinfo {year}
  {1999})}\BibitemShut {NoStop}%
\bibitem [{\citenamefont {Boninsegni}\ \emph {et~al.}(2006)\citenamefont
  {Boninsegni}, \citenamefont {Prokof'ev},\ and\ \citenamefont
  {Svistunov}}]{PhysRevE.74.036701}%
  \BibitemOpen
  \bibfield  {author} {\bibinfo {author} {\bibfnamefont {M.}~\bibnamefont
  {Boninsegni}}, \bibinfo {author} {\bibfnamefont {N.~V.}\ \bibnamefont
  {Prokof'ev}}, \ and\ \bibinfo {author} {\bibfnamefont {B.~V.}\ \bibnamefont
  {Svistunov}},\ }\href {\doibase 10.1103/PhysRevE.74.036701} {\bibfield
  {journal} {\bibinfo  {journal} {Phys. Rev. E}\ }\textbf {\bibinfo {volume}
  {74}},\ \bibinfo {pages} {036701} (\bibinfo {year} {2006})}\BibitemShut
  {NoStop}%
\bibitem [{\citenamefont {Burovski}\ \emph {et~al.}(2008)\citenamefont
  {Burovski}, \citenamefont {Kozik}, \citenamefont {Prokof'ev}, \citenamefont
  {Svistunov},\ and\ \citenamefont {Troyer}}]{PhysRevLett.101.090402}%
  \BibitemOpen
  \bibfield  {author} {\bibinfo {author} {\bibfnamefont {E.}~\bibnamefont
  {Burovski}}, \bibinfo {author} {\bibfnamefont {E.}~\bibnamefont {Kozik}},
  \bibinfo {author} {\bibfnamefont {N.}~\bibnamefont {Prokof'ev}}, \bibinfo
  {author} {\bibfnamefont {B.}~\bibnamefont {Svistunov}}, \ and\ \bibinfo
  {author} {\bibfnamefont {M.}~\bibnamefont {Troyer}},\ }\href {\doibase
  10.1103/PhysRevLett.101.090402} {\bibfield  {journal} {\bibinfo  {journal}
  {Phys. Rev. Lett.}\ }\textbf {\bibinfo {volume} {101}},\ \bibinfo {pages}
  {090402} (\bibinfo {year} {2008})}\BibitemShut {NoStop}%
\bibitem [{\citenamefont {Assaad}\ and\ \citenamefont
  {Lang}(2007)}]{PhysRevB.76.035116}%
  \BibitemOpen
  \bibfield  {author} {\bibinfo {author} {\bibfnamefont {F.~F.}\ \bibnamefont
  {Assaad}}\ and\ \bibinfo {author} {\bibfnamefont {T.~C.}\ \bibnamefont
  {Lang}},\ }\href {\doibase 10.1103/PhysRevB.76.035116} {\bibfield  {journal}
  {\bibinfo  {journal} {Phys. Rev. B}\ }\textbf {\bibinfo {volume} {76}},\
  \bibinfo {pages} {035116} (\bibinfo {year} {2007})}\BibitemShut {NoStop}%
\bibitem [{\citenamefont {C\`e}\ \emph {et~al.}(2017)\citenamefont {C\`e},
  \citenamefont {Giusti},\ and\ \citenamefont {Schaefer}}]{PhysRevD.95.034503}%
  \BibitemOpen
  \bibfield  {author} {\bibinfo {author} {\bibfnamefont {M.}~\bibnamefont
  {C\`e}}, \bibinfo {author} {\bibfnamefont {L.}~\bibnamefont {Giusti}}, \ and\
  \bibinfo {author} {\bibfnamefont {S.}~\bibnamefont {Schaefer}},\ }\href
  {\doibase 10.1103/PhysRevD.95.034503} {\bibfield  {journal} {\bibinfo
  {journal} {Phys. Rev. D}\ }\textbf {\bibinfo {volume} {95}},\ \bibinfo
  {pages} {034503} (\bibinfo {year} {2017})}\BibitemShut {NoStop}%
\bibitem [{\citenamefont {Ayyar}\ and\ \citenamefont
  {Chandrasekharan}(2016)}]{PhysRevD.93.081701}%
  \BibitemOpen
  \bibfield  {author} {\bibinfo {author} {\bibfnamefont {V.}~\bibnamefont
  {Ayyar}}\ and\ \bibinfo {author} {\bibfnamefont {S.}~\bibnamefont
  {Chandrasekharan}},\ }\href {\doibase 10.1103/PhysRevD.93.081701} {\bibfield
  {journal} {\bibinfo  {journal} {Phys. Rev. D}\ }\textbf {\bibinfo {volume}
  {93}},\ \bibinfo {pages} {081701} (\bibinfo {year} {2016})}\BibitemShut
  {NoStop}%
\bibitem [{\citenamefont {Kaul}\ \emph {et~al.}(2013)\citenamefont {Kaul},
  \citenamefont {Melko},\ and\ \citenamefont {Sandvik}}]{RevCondMat}%
  \BibitemOpen
  \bibfield  {author} {\bibinfo {author} {\bibfnamefont {R.~K.}\ \bibnamefont
  {Kaul}}, \bibinfo {author} {\bibfnamefont {R.~G.}\ \bibnamefont {Melko}}, \
  and\ \bibinfo {author} {\bibfnamefont {A.~W.}\ \bibnamefont {Sandvik}},\
  }\href {\doibase 10.1146/annurev-conmatphys-030212-184215} {\bibfield
  {journal} {\bibinfo  {journal} {Annual Review of Condensed Matter Physics}\
  }\textbf {\bibinfo {volume} {4}},\ \bibinfo {pages} {179} (\bibinfo {year}
  {2013})}\BibitemShut {NoStop}%
\bibitem [{\citenamefont {Assaad}\ and\ \citenamefont
  {Grover}(2016)}]{PhysRevX.6.041049}%
  \BibitemOpen
  \bibfield  {author} {\bibinfo {author} {\bibfnamefont {F.~F.}\ \bibnamefont
  {Assaad}}\ and\ \bibinfo {author} {\bibfnamefont {T.}~\bibnamefont
  {Grover}},\ }\href {\doibase 10.1103/PhysRevX.6.041049} {\bibfield  {journal}
  {\bibinfo  {journal} {Phys. Rev. X}\ }\textbf {\bibinfo {volume} {6}},\
  \bibinfo {pages} {041049} (\bibinfo {year} {2016})}\BibitemShut {NoStop}%
\bibitem [{\citenamefont {Wang}\ \emph {et~al.}(2016)\citenamefont {Wang},
  \citenamefont {Liu},\ and\ \citenamefont {Troyer}}]{PhysRevB.93.155117}%
  \BibitemOpen
  \bibfield  {author} {\bibinfo {author} {\bibfnamefont {L.}~\bibnamefont
  {Wang}}, \bibinfo {author} {\bibfnamefont {Y.-H.}\ \bibnamefont {Liu}}, \
  and\ \bibinfo {author} {\bibfnamefont {M.}~\bibnamefont {Troyer}},\ }\href
  {\doibase 10.1103/PhysRevB.93.155117} {\bibfield  {journal} {\bibinfo
  {journal} {Phys. Rev. B}\ }\textbf {\bibinfo {volume} {93}},\ \bibinfo
  {pages} {155117} (\bibinfo {year} {2016})}\BibitemShut {NoStop}%
\bibitem [{\citenamefont {Susskind}(1977)}]{PhysRevD.16.3031}%
  \BibitemOpen
  \bibfield  {author} {\bibinfo {author} {\bibfnamefont {L.}~\bibnamefont
  {Susskind}},\ }\href {\doibase 10.1103/PhysRevD.16.3031} {\bibfield
  {journal} {\bibinfo  {journal} {Phys. Rev. D}\ }\textbf {\bibinfo {volume}
  {16}},\ \bibinfo {pages} {3031} (\bibinfo {year} {1977})}\BibitemShut
  {NoStop}%
\bibitem [{\citenamefont {Huffman}\ and\ \citenamefont
  {Chandrasekharan}(2014)}]{Huffman:2013mla}%
  \BibitemOpen
  \bibfield  {author} {\bibinfo {author} {\bibfnamefont {E.~F.}\ \bibnamefont
  {Huffman}}\ and\ \bibinfo {author} {\bibfnamefont {S.}~\bibnamefont
  {Chandrasekharan}},\ }\href {\doibase 10.1103/PhysRevB.89.111101} {\bibfield
  {journal} {\bibinfo  {journal} {Phys. Rev.}\ }\textbf {\bibinfo {volume}
  {B89}},\ \bibinfo {pages} {111101} (\bibinfo {year} {2014})},\ \Eprint
  {http://arxiv.org/abs/1311.0034} {arXiv:1311.0034 [cond-mat.str-el]}
  \BibitemShut {NoStop}%
%%CITATION = ARXIV:1311.0034;%%
\bibitem [{\citenamefont {Li}\ \emph {et~al.}(2015{\natexlab{b}})\citenamefont
  {Li}, \citenamefont {Jiang},\ and\ \citenamefont {Yao}}]{PhysRevB.91.241117}%
  \BibitemOpen
  \bibfield  {author} {\bibinfo {author} {\bibfnamefont {Z.-X.}\ \bibnamefont
  {Li}}, \bibinfo {author} {\bibfnamefont {Y.-F.}\ \bibnamefont {Jiang}}, \
  and\ \bibinfo {author} {\bibfnamefont {H.}~\bibnamefont {Yao}},\ }\href
  {\doibase 10.1103/PhysRevB.91.241117} {\bibfield  {journal} {\bibinfo
  {journal} {Phys. Rev. B}\ }\textbf {\bibinfo {volume} {91}},\ \bibinfo
  {pages} {241117} (\bibinfo {year} {2015}{\natexlab{b}})}\BibitemShut
  {NoStop}%
\bibitem [{\citenamefont {Agranovich}\ and\ \citenamefont
  {Maradudin}(1992)}]{inbook}%
  \BibitemOpen
  \bibfield  {author} {\bibinfo {author} {\bibfnamefont {V.}~\bibnamefont
  {Agranovich}}\ and\ \bibinfo {author} {\bibfnamefont {A.~A.}\ \bibnamefont
  {Maradudin}},\ }\enquote {\bibinfo {title} {Electronic phase transitions},}\
  \ (\bibinfo  {publisher} {Elsevier Science Publishers B. V.},\ \bibinfo
  {year} {1992})\ Chap.~\bibinfo {chapter} {4}, pp.\ \bibinfo {pages}
  {177--235}\BibitemShut {NoStop}%
\bibitem [{Sup()}]{Suppmat}%
  \BibitemOpen
  \href@noop {} {}\bibinfo {note} {For further details we refer the reader to
  the supplementary materials attached}\BibitemShut {NoStop}%
\bibitem [{\citenamefont {Rosa}\ \emph {et~al.}(2001)\citenamefont {Rosa},
  \citenamefont {Vitale},\ and\ \citenamefont
  {Wetterich}}]{PhysRevLett.86.958}%
  \BibitemOpen
  \bibfield  {author} {\bibinfo {author} {\bibfnamefont {L.}~\bibnamefont
  {Rosa}}, \bibinfo {author} {\bibfnamefont {P.}~\bibnamefont {Vitale}}, \ and\
  \bibinfo {author} {\bibfnamefont {C.}~\bibnamefont {Wetterich}},\ }\href
  {\doibase 10.1103/PhysRevLett.86.958} {\bibfield  {journal} {\bibinfo
  {journal} {Phys. Rev. Lett.}\ }\textbf {\bibinfo {volume} {86}},\ \bibinfo
  {pages} {958} (\bibinfo {year} {2001})}\BibitemShut {NoStop}%
\bibitem [{\citenamefont {Mihaila}\ \emph {et~al.}(2017)\citenamefont
  {Mihaila}, \citenamefont {Zerf}, \citenamefont {Ihrig}, \citenamefont
  {Herbut},\ and\ \citenamefont {Scherer}}]{Mihaila:2017ble}%
  \BibitemOpen
  \bibfield  {author} {\bibinfo {author} {\bibfnamefont {L.~N.}\ \bibnamefont
  {Mihaila}}, \bibinfo {author} {\bibfnamefont {N.}~\bibnamefont {Zerf}},
  \bibinfo {author} {\bibfnamefont {B.}~\bibnamefont {Ihrig}}, \bibinfo
  {author} {\bibfnamefont {I.~F.}\ \bibnamefont {Herbut}}, \ and\ \bibinfo
  {author} {\bibfnamefont {M.~M.}\ \bibnamefont {Scherer}},\ }\href@noop {} {\
  (\bibinfo {year} {2017})},\ \Eprint {http://arxiv.org/abs/1703.08801}
  {arXiv:1703.08801 [cond-mat.str-el]} \BibitemShut {NoStop}%
%%CITATION = ARXIV:1703.08801;%%
\bibitem [{\citenamefont {Cardy}(1988)}]{Cardy:1988ag}%
  \BibitemOpen
  \bibinfo {editor} {\bibfnamefont {J.~L.}\ \bibnamefont {Cardy}},\ ed.,\
  \href@noop {} {\emph {\bibinfo {title} {{FINITE SIZE SCALING}}}}\ (\bibinfo
  {year} {1988})\BibitemShut {NoStop}%
%%CITATION = INSPIRE-272420;%%
\bibitem [{\citenamefont {Campostrini}\ \emph {et~al.}(2014)\citenamefont
  {Campostrini}, \citenamefont {Pelissetto},\ and\ \citenamefont
  {Vicari}}]{PhysRevB.89.094516}%
  \BibitemOpen
  \bibfield  {author} {\bibinfo {author} {\bibfnamefont {M.}~\bibnamefont
  {Campostrini}}, \bibinfo {author} {\bibfnamefont {A.}~\bibnamefont
  {Pelissetto}}, \ and\ \bibinfo {author} {\bibfnamefont {E.}~\bibnamefont
  {Vicari}},\ }\href {\doibase 10.1103/PhysRevB.89.094516} {\bibfield
  {journal} {\bibinfo  {journal} {Phys. Rev. B}\ }\textbf {\bibinfo {volume}
  {89}},\ \bibinfo {pages} {094516} (\bibinfo {year} {2014})}\BibitemShut
  {NoStop}%
\end{thebibliography}%

\onecolumngrid

\newpage
\appendix*

\begin{center}
{\Large \bf Supplementary Material}
\end{center}

Here we discuss some of the missing details that a reader may want to understand.

\section*{Details of the Updates}

In Section III of the paper we discuss how we compute the ratio $R$, defined by
\begin{equation}
R \ =\ \det(\mathbbm{1}_N + M_B M_T')/\det(\mathbbm{1}_N + M_B M_T) \ =\ \det\left(\mathbbm{1}_N + G_B\Delta \right),
\label{ratR}
\end{equation}
where $G_B$ is $G_B = \left(\mathbbm{1}_N+M_B M_T\right)^{-1} M_B M_T$ and $\Delta = \left( M_T^{-1} M_T' - \mathbbm{1}_N\right)$. While this $R$ is for any ratio of an update configuration weight to its background configuration, less generically we can define $R_{\rm curr}$ to be such a ratio for a current configuration within a block update, and $R_{\rm new}$ to be such a ratio for a proposed update. The weight ratio that we need is then found from $R_{\rm new}/R_{\rm curr}$. Since we usually have already found $R_{\rm curr}$ from a previous update proposal, the new update calculation usually consists of only one use of (\ref{ratR}).

A naive computation of the matrices $M_B$, $M_T$ and $M_T'$ can be numerically unstable as is well known. Note that each of these matrices are constructed as a product of block matrices $B_{x,d}$. In a typical auxiliary field Monte Carlo method the product is accomplished using the singular value decomposition (SVD) of individual matrices $B_{x,d}$ that are contained in these matrices. This is time consuming and we would like to avoid it as much as possible.

In our case since each $B_{x,d}$ is only non-trivial in a $2\times 2$ block we can multiply a bunch of them at a time without worrying about SVDs. We call each such bunch as a partial product $M_i$. However the remaining product will still need SVDs in principle. To accommodate this we package the information needed in $R$ into the matrices $G_B$ and $\Delta$. Then there are three main numerical instabilities that we have to deal with: (1) computing $G_B$ and then updating it when we move on to a different time-slice, (2) updating $G_B$ between block-updates within the same time-slice, and (3) updating $\Delta$ for each configuration update and ensuring a stable determinant. We will discuss how we accomplish these in some detail below.

First note that given two matrices, $M_1$ and $M_2$, we have the identity
\begin{equation}
\begin{aligned}
\left( \mathbbm{1}+M_1  M_2\right)^{-1} 
= &\\ 
\left(\mathbbm{1}+M_2\right)^{-1} & \left(\left(\mathbbm{1}+M_1\right)^{-1} \left(\mathbbm{1}+M_2\right)^{-1}\right.
\left. + \left(\mathbbm{1}+M_1\right)^{-1}M_1 M_2\left(\mathbbm{1}+M_2\right)^{-1}\right)^{-1} \left(\mathbbm{1}+M_1\right)^{-1}.
\label{inv}
\end{aligned}
\end{equation}
Further it is convenient that $\left( \mathbbm{1}+ M \right)^{-1} M =  \mathbbm{1} - (\mathbbm{1}+M)^{-1}$. This means we can build $G_B$ from partial versions labeled as $G_T=(1+M_T)^{-1}M_T$ associated with each time slice $T$. The matrix $G_T$ in turn is obtained by combining the partial $G_i= (1+M_i)^{-1}M_i$ within a time slice, where the matrices $M_i$ are the partial products we explained above. We can do this efficiently using the idea of fermion bags. For each fermion bag in a given time-slice, we first construct $G_f$ taking into account the corresponding $G_i$'s that belong to a specific fermion bag. Thus each $G_f$ is a matrix with $f$ rows and $f$ columns corresponding to the fermion bag sites. We can then combine the $G_f$ matrices into a $G_T$ matrix, which has distinct blocks according to the fermion bags. Thus, while $G_B$ is an $N\times N$ matrix, we can use the idea of fermion bags along with the identity (\ref{inv}) to reduce the number of $O\left( N^3 \right)$ operations. While this does not reduce the scaling of the algorithm, it does significantly reduce the prefactor.

As we build $G_B$, the partial forms of $G_B$ are already stored either in computer memory or on the hard disk. The stored partial forms allow us to make fast updates to $G_B$ when we move sequentially through the time-slices and keep the linear $\beta$ scaling. More details on how the storage scheme works can be found in \cite{Wang:2015rga}. One wrinkle is that at times the $O_2$ matrix can cause a singularity so that certain types of partial $G_B$ matrices do not exist. If that happens we simply wait to combine the $O_2$ matrix with the others until the final combination to form $G_B$.

A second important update to $G_B$ occurs when we are ready to update a new block without changing time-slices. The $G_B$ must be updated according to the \textit{super-bag} $S$ of the current block. Here we use the Woodbury identity to make the update only of order $O\left(sN^2\right)$. Assuming $G_B = \left(\mathbbm{1} + M_B M_T\right)M_B M_T$ before the update,
\begin{equation}
\begin{aligned}
G_B' &=\left(\mathbbm{1}+M_B M +  M_B M \left(M_T^{-1} M_T' - \mathbbm{1}\right)\right)^{-1}\\
&=\mathbbm{1}-G_B - \left[G_B\right]_{N\times s} \left(\left[ \mathbbm{1}-\mathcal{G}_T + G_B - 2 \right(\mathbbm{1}-\mathcal{G}_T\left) G_B \right]_{s\times s}\right)^{-1}\left[\left(\mathbbm{1} -2\mathcal{G}_T \right)\right]_{s\times s} \left[\left(\mathbbm{1}-G_B\right)\right]_{s\times N},
\end{aligned}
\end{equation}
where $M_T$ and ${M_T}'$ are matrix products in timeslice $T$ for the configurations that go with $G_B$ and $G_B'$, respectively, and $\mathcal{G}_T = \left(\mathbbm{1}+M_T^{-1} {M_T}'\right)^{-1}$. The symbol $\left[\;\right]_{s\times s}$ means only the rows and columns belonging to the super-bag $S$ are used, with $\left[\;\right]_{N\times s}$ and $\left[\;\right]_{s\times N}$ forming matrices from columns belonging to $S$ and rows belonging to $S$, respectively.

Finally, we update the matrices $M_T'$ that are found in $\Delta$ often. For ease of computation and to ensure stability, the quantity we update is actually $\square = G_B {M_T}^{-1} {M_T}'$, and the determinant we calculate is
\begin{equation}
R = \det\left(\left[\mathbbm{1}-G_B + \square\right]_{s\times s}\right) = \left|\det\left(\left[\left(\mathbbm{1}-G_B\right)\mathcal{Q}^T + \mathcal{R}\right]_{s\times s} \right) \right|,
\label{qrrat}
\end{equation}
where we are using the $RQ$ factorization of $\square$ into an upper triangular matrix $\mathcal{R}$ and an orthogonal matrix $\mathcal{Q}$, as in \cite{inbook}. Only the ${M_T}'$ matrices have to be updated each time, so we store an $RQ$ factorization of the $G_B {M_T}^{-1}$ product for the block update.

\section*{Algorithm Performance}

\begin{figure}
\minipage{0.48\textwidth}
\includegraphics[width=\linewidth]{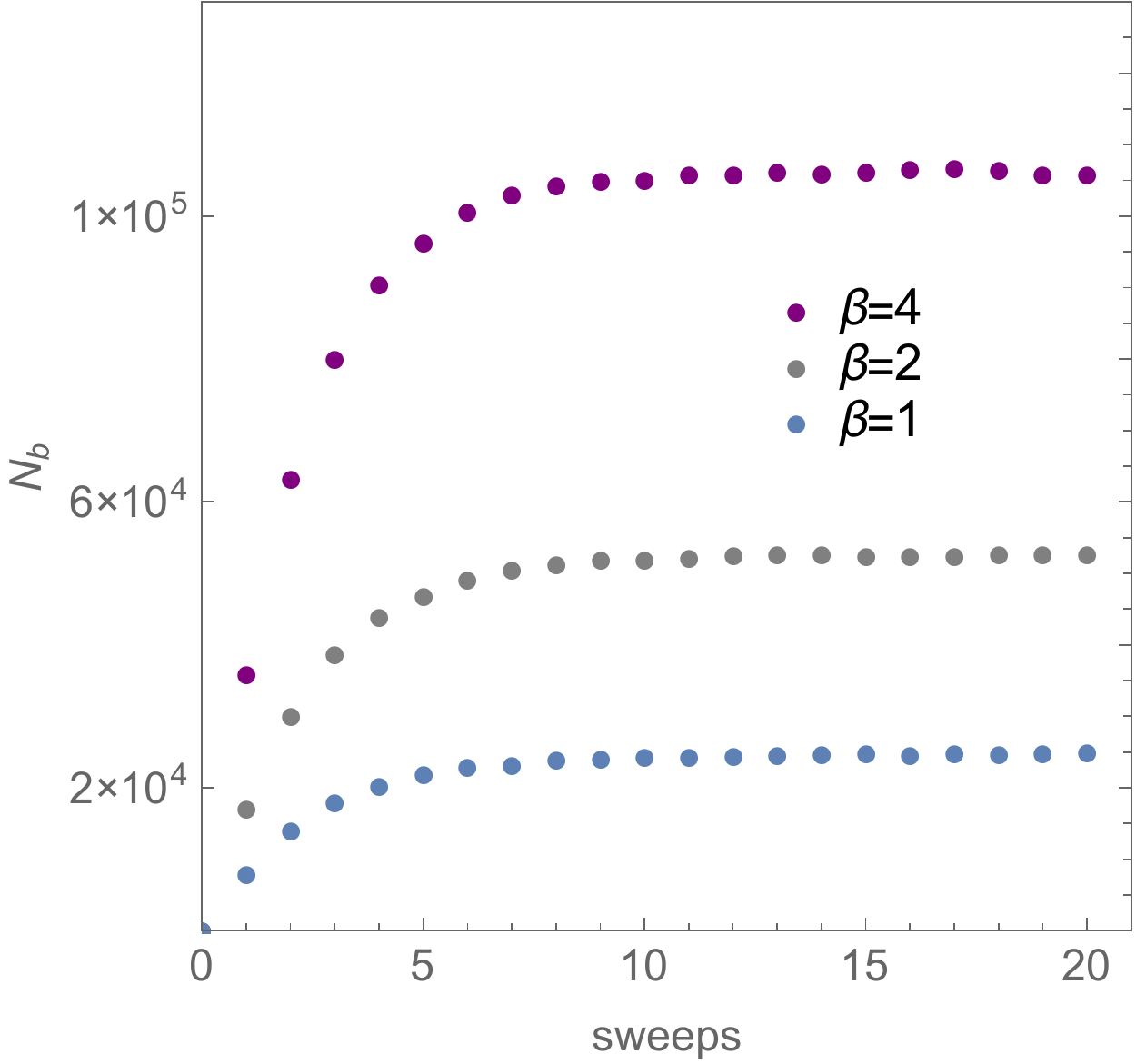}
\caption{Plot showing equilibration of the bond number for $V/t=1.304$, $L=100$ configurations with $\beta=1,2,4$ as a function of sweeps.}
\label{smallbeta}
\endminipage\hfill
\minipage{0.48\textwidth}
\includegraphics[width=\linewidth]{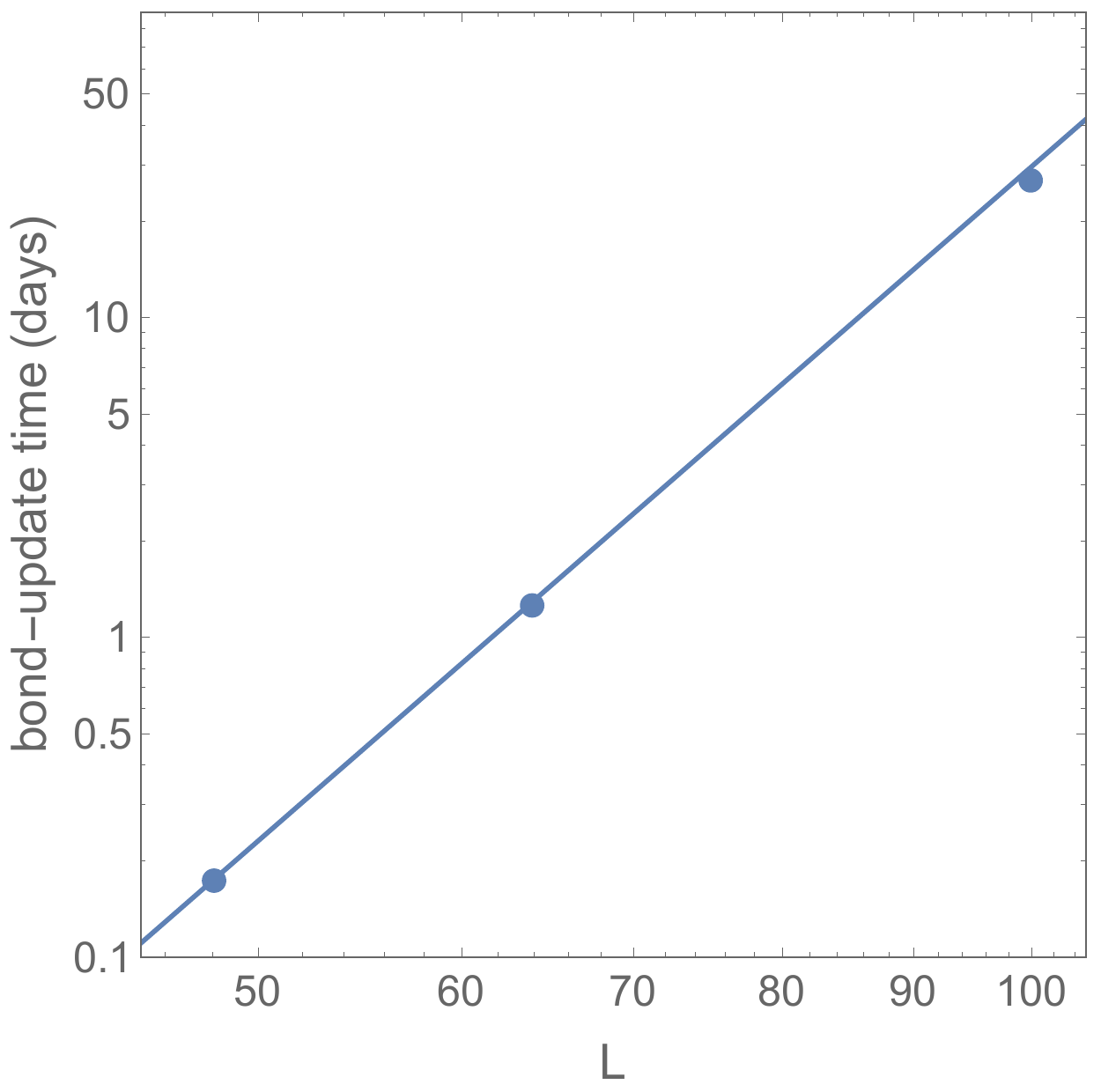}
\caption{Plot showing the time to complete a single \textit{bond-update} (in days) for $L=48,64,100$ with $\beta=L$ at $V/t=1.304$. The solid line is a plot of $\tau = 3 \times 10^{-13} L^7$.}
\label{scales}
\endminipage
\end{figure}

As mentioned in Section III of the paper, we can easily equilibrate even $L=100$ lattices for small $\beta$ values. Fig. \ref{smallbeta} shows some equilibrations for the small $\beta$ values of $1,2$ amd $4$ at $V/t=1.304$ (this is currently also shown in the inset of Fig.4 of the paper). In Fig. \ref{scales}, we confirm the $O(\beta N^3)$ scaling of time for a complete \textit{bond-update}. In particular we plot the bond update time $\tau_b$ (in days) as a function of $L$ for three different lattice sizes at the coupling $V/t=1.304$ close to the critical point. Since $\beta=L$ we expect a scaling of $O(L^7)$. As expected the solid line in the figure, which is the plot of $\tau_b = 3 \times 10^{-13} L^7$, roughly passes through all the points.

\begin{figure}[ht]
\minipage{0.3\textwidth}
  \includegraphics[width=\linewidth]{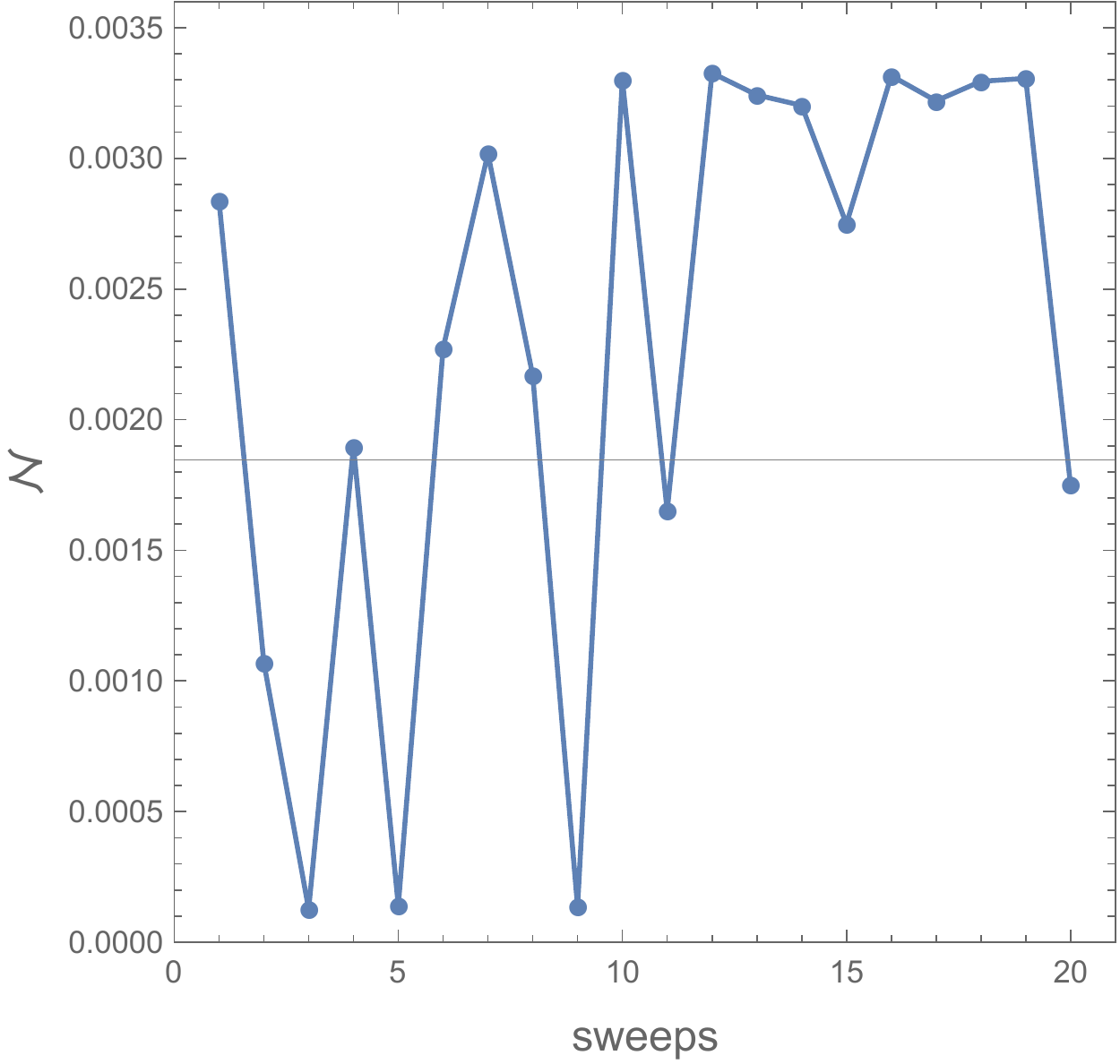}
\endminipage\hfill
\minipage{0.3\textwidth}
  \includegraphics[width=\linewidth]{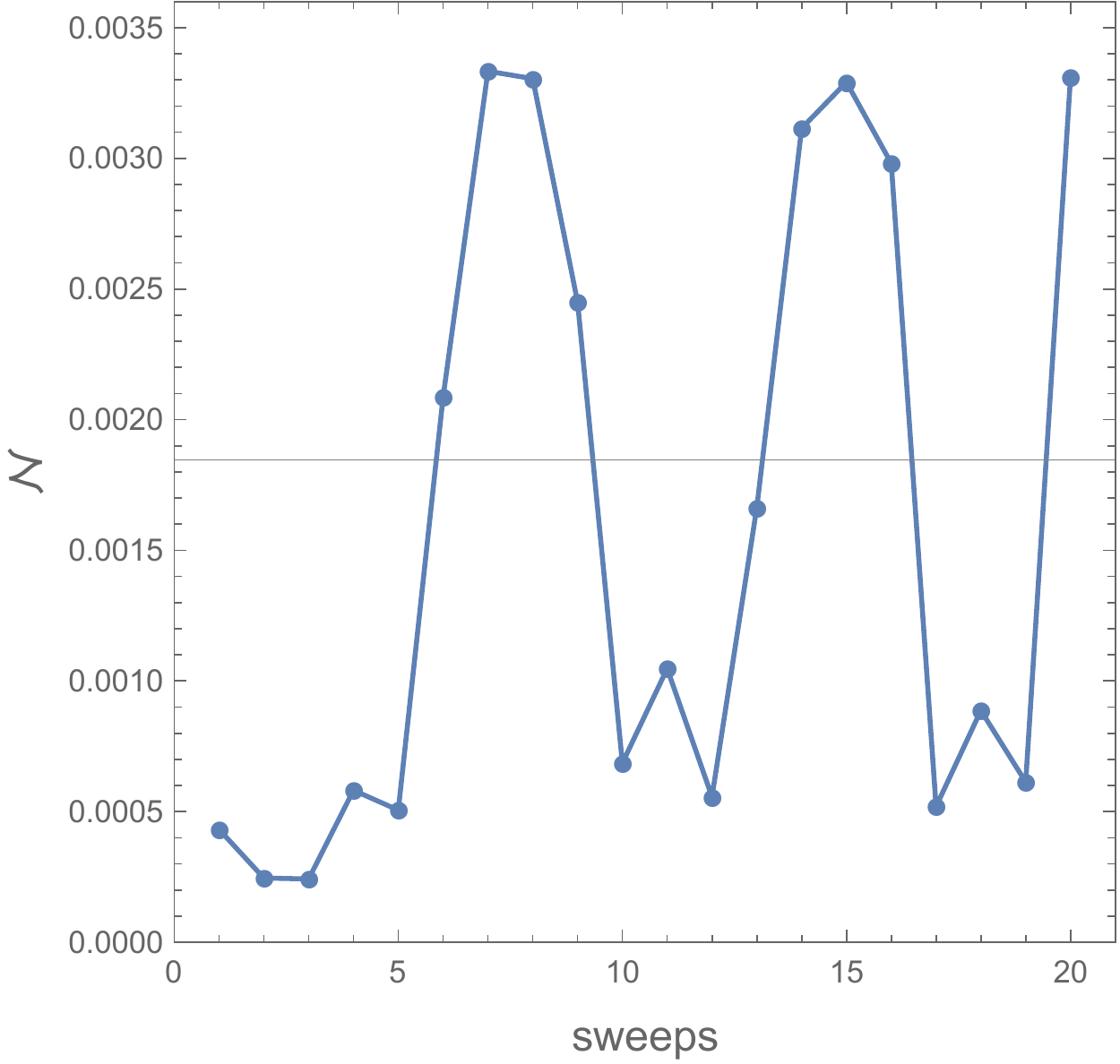}
\endminipage\hfill
\minipage{0.3\textwidth}%
  \includegraphics[width=\linewidth]{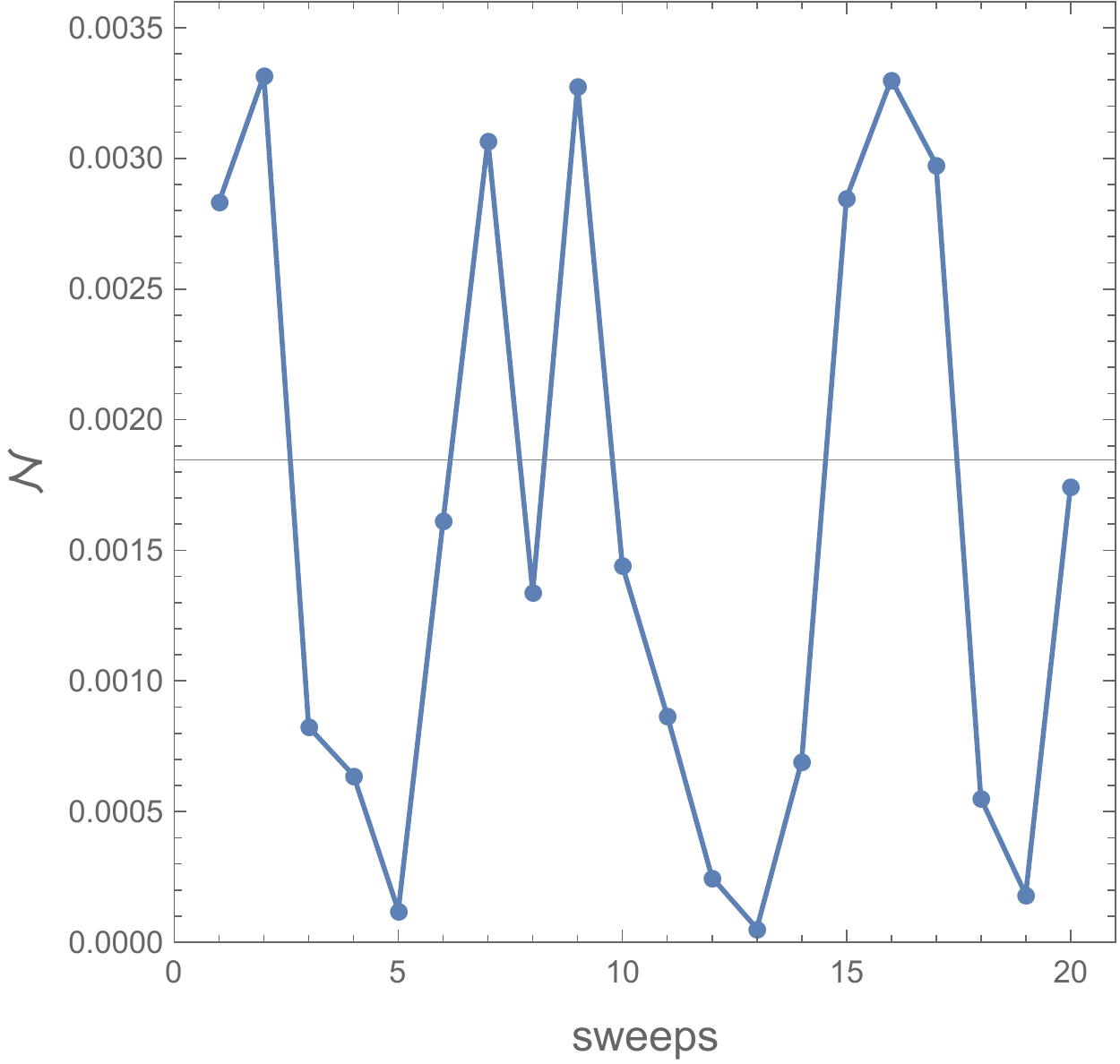}
\endminipage
 \caption{Measurements of $\mathcal{N}$ taken from different threads starting from the same equilibrated configuration for $V=1.304$ at $L=48$. The first two measurements have been discarded. The full average, $.0019$, is given by the gray line in each plot. The first plot has an average of $.0023$, the second an average of $.0016$, and the third an average of $.0014$.}\label{configurations}
\end{figure}

In order to compute our observable we need to generate a large number of statistically independent configurations. Through the Open Science Grid, we have access to several hundreds of CPU cores at a time. Thus, we can typically run about $1000$ independent threads of our algorithm. On small lattices we start $1000$ independent runs from a configuration without any bonds. We then wait for equilibration and collect about $20$ sweeps of data from each thread thus generating statistics of about $20^4$ configurations.

On larger lattices we equilibrate $10$ independent configurations and copy each configuration on $100$ cores. Thus we start the $1000$ cores with equilibrated configurations but many of which are completely correlated. In Fig.~\ref{configurations} we plot the Monte Carlo fluctuations of three such threads starting from the same equilibrated configuration with different random number sequences. We note that the observable ${\cal N}$ defined in Eq.(\ref{nobs}) of the paper seems to become decorrelated within a few sweeps. Hence, we again can generate $20$ sweeps of data on each of the $1000$ threads. We compute averages after throwing away the first few sweeps.

\section*{Tests of the Algorithm}

We have tested our algorithm in multiple ways. One of the main steps in the algorithm involves computing ratios $R$ of configuration weights $\Omega_n([x,d,t];t_0)$ defined by the trace
\begin{equation}
\Omega_n([x,d,t];t_0) = {\rm Tr}\left[H_{x_k, d_k} ...C_n... H_{x_2,d_2} H_{x_1, d_1}\right].
\end{equation}
While this weight can be computed using the BSS formula that is used in the our algorithm (see Eq.~(\ref{ratR})),
\begin{equation}
\Omega_n\left([x,t,b];t_0\right)=\det\left(\mathbbm{1}_N+B_{x_k,d_k}...O_n...B_{x_2,d_2} B_{x_1,d_1}\right),
\end{equation}
this approach can be unstable. Fortunately, it can also be computed as the determinant of a $2k\times 2k$ antisymmetric matrix, where $k$ is the number of bond insertions. Although the calculation of this determinant is more time consuming, it is stable. We have computed $R$ by both these methods on small lattices and confirmed that they agree to very high accuracy.

We have also compared the results for the observable $\langle C\rangle$ obtained from the Monte Carlo algorithm against exact calculations. Table \ref{tablecheck} summarizes the results on 2$\times$2 and 4$\times$4 lattices at couplings $V=1.200,1.304$ for inverse temperature values of $\beta=1.0,2.0,4.0,8.0$. We have also verified that the value the reweighting factor $f$ does not affect the observable. We show our results for $f=10.0$ and $50.0$. The table shows that the Monte Carlo results agree with the exact calculations within errors as expected. Additionally, similar tests were performed at $V=1.304$ for $f=1.0,2.0,$ and $f=20.0$, and again the exact results were within the errors of the Monte Carlo results.

\begin{table}[t]
\begin{tabular*}{.47\linewidth}{@{\extracolsep{\fill}}|c|c|c|c|c|}
\hline
\multicolumn{5}{|c|}{\textbf{2$\times$2 Lattice, V=1.200}} \\
\hline
& $\beta=1.0$ & $\beta=2.0$ & $\beta=4.0$ & $\beta=8.0$  \\
\hline
\hline
MC (f=10) & 0.09838(16) & 0.14265(15) & 0.15091(9) & 0.15105(7) \\
\hline
\hline
MC (f=50) & 0.09872(23) & 0.14242(19) & 0.15082(13) & 0.15107(11) \\
\hline
\hline
Exact & 0.098550... & 0.142590... & 0.150801... & 0.150939... 
\\
\hline
\end{tabular*}
\begin{tabular*}{.47\linewidth}{@{\extracolsep{\fill}}|c|c|c|c|c|}
\hline
\multicolumn{5}{|c|}{\textbf{2$\times$2 Lattice, V=1.304}} \\
\hline
& $\beta=1.0$ & $\beta=2.0$ & $\beta=4.0$ & $\beta=8.0$  \\
\hline
\hline
MC (f=10) & 0.10290(17) & 0.14562(13) & 0.15286(9) & 0.15320(7) \\
\hline
\hline
MC (f=50) & 0.10261(25) & 0.14557(18) & 0.15287(13) & 0.15305(11) \\
\hline
\hline
Exact & 0.102948... & 0.145738... & 0.152973... & 0.153078... 
\\
\hline
\end{tabular*}
\begin{tabular*}{.47\linewidth}{@{\extracolsep{\fill}}|c|c|c|c|c|}
\hline
\multicolumn{5}{|c|}{\textbf{4$\times$4 Lattice, V=1.200}} \\
\hline
& $\beta=1.0$ & $\beta=2.0$ & $\beta=4.0$ & $\beta=8.0$  \\
\hline
\hline
MC (f=10) & 0.03117(8) & 0.05955(12) & 0.07781(15) & 0.07997(14) \\
\hline
\hline
MC (f=50) & 0.03127(8) & 0.05948(13) & 0.07780(16) & 0.07994(16) \\
\hline
\hline
Exact & 0.031285... & 0.059458... & 0.077769... & 0.080009... 
\\
\hline
\end{tabular*}
\begin{tabular*}{.47\linewidth}{@{\extracolsep{\fill}}|c|c|c|c|c|}
\hline
\multicolumn{5}{|c|}{\textbf{4$\times$4 Lattice, V=1.304}} \\
\hline
& $\beta=1.0$ & $\beta=2.0$ & $\beta=4.0$ & $\beta=8.0$  \\
\hline
\hline
MC (f=10) & 0.03819(10) & 0.07272(15) & 0.08959(17) & 0.09060(16) \\
\hline
\hline
MC (f=50) & 0.03798(13) & 0.07293(16) & 0.08934(18) & 0.09087(18) \\
\hline
\hline
Exact & 0.038105... & 0.072760... & 0.089511... & 0.090672... 
\\
\hline
\end{tabular*}
\caption{Monte Carlo measurements (MC) and exact calculation values (Exact) on small lattices for different parameter combinations.}
\label{tablecheck}
\end{table}

\end{document}